\documentclass[%
 reprint,
 superscriptaddress,
 twocolumn, 
 amsmath,amssymb,
 aps,
]{revtex4-2}
\usepackage{todonotes}
\usepackage{breakurl}
\usepackage[colorlinks = true,
            linkcolor = blue,
            urlcolor  = blue,
            citecolor = blue,
            anchorcolor = blue]{hyperref}
\usepackage[normalem]{ulem}
\usepackage{graphicx}% Include figure files
\usepackage{dcolumn}% Align table columns on decimal point
\usepackage{bm}% bold math
\usepackage{braket}
\usepackage{glossaries}
\usepackage{subfigure}
\newacronym{CDW}{CDW}{charge-density-wave}
\newacronym{QMC}{QMC}{quantum monte carlo}
\newacronym{DFPT}{DFPT}{density functional perturbation theory}
\newacronym{DFT}{DFT}{density functional theory}

\usepackage{ulem}

\makeatletter
\newsavebox{\@brx}
\newcommand{\llangle}[1][]{\savebox{\@brx}{\(\m@th{#1\langle}\)}%
  \mathopen{\copy\@brx\kern-0.5\wd\@brx\usebox{\@brx}}}
\newcommand{\rrangle}[1][]{\savebox{\@brx}{\(\m@th{#1\rangle}\)}%
  \mathclose{\copy\@brx\kern-0.5\wd\@brx\usebox{\@brx}}}
\makeatother

\newcommand{\COMMENTED}[1]{}

\begin{document}

\preprint{}

\title{Interaction-Driven Ferrimagnetic Stripes in the Extended Hubbard Model}

\author{Chunhan Feng}
\affiliation{Center for Computational Quantum Physics, Flatiron Institute, 162 5th Avenue, New York, New York, USA}
\affiliation{Max Planck Institute for the Physics of Complex Systems, Nöthnitzer Straße 38, 01187 Dresden, Germany}
\author{ Miguel A. Morales}
\affiliation{Center for Computational Quantum Physics, Flatiron Institute, 162 5th Avenue, New York, New York, USA}
\author{Shiwei Zhang}
\affiliation{Center for Computational Quantum Physics, Flatiron Institute, 162 5th Avenue, New York, New York, USA}

\date{\today}

\begin{abstract}
Long-range interactions can qualitatively reorganize 
correlated-electron ground states. In the square-lattice Hubbard model,  
on-site repulsion produces antiferromagnetic spin and charge stripes upon doping. We show that including a nearest-neighbor repulsion $V$  
can dramatically
alter this behavior. Using auxiliary-field quantum Monte Carlo and density matrix renormalization group methods, we find that, above a critical ratio $V/U$ ($\sim 0.25$), 
the system develops a modulated ferrimagnetic order intertwined with checkerboard charge-density-wave.  
Inside the ferrimagnetic domains, spin density alternates between positive (or negative) and nearly zero values. When the total spin is fixed to zero, positive and negative domains alternate in space; when spins are unconstrained, a ferrimagnetic state emerges with finite magnetization.
Including a next-nearest-neighbor hopping $t'$ changes the modulation wavelength but leaves the order 
robust.
Our results demonstrate that even short-range nonlocal interactions can stabilize qualitatively new magnetic textures, with implications for cuprate materials and programmable quantum simulators.

\end{abstract}

\maketitle

%%%%%%%%%%%%%%%%%%%%%%%%%%%%%%%%%%%%%%%%%%%%%%%%%%%%%%%%%%%%%%%%%%%%%%%%%%%%%%%%%%%%%%%%%
%%%%%%%%%%%%%%%%%%%%%%%%%%%%%%%%%%%%%%%%%%%%%%%%%%%%%%%%%%%%%%%%%%%%%%%%%%%%%%%%%%%%%%%%%
\noindent
\textit{Introduction}.~
The Hubbard model \cite{Hubbard1963} occupies a central role in quantum physics, having long served as the minimal framework for understanding strong electron correlations. 
Despite its simplicity, it captures emergent behavior ranging from antiferromagnetism and stripe correlations to pseudogap physics and unconventional superconductivity \cite{Kivelson2003,Proust2019,Arovas2022,Qin2020,Wietek2021,Xu2023,Simkovic2024}.
In real materials, however, electrons interact via the long-range Coulomb force,
suggesting that it is crucial to understand the role of 
nonlocal terms beyond the onsite repulsion 
$U$.
Indeed, extensions of the simplest Hubbard model have shown that additional
terms can qualitatively reshape the ground-state landscape. For example,
next-nearest-neighbor hopping $t^\prime$
induces pronounced particle–hole asymmetry in stripe and pairing tendencies \cite{Xu2023},
demonstrating the sensitivity of electronic order to nonlocal energy scales.

A nearest-neighbor Coulomb repulsion $V$ is the simplest and most physically relevant nonlocal interaction. While the local $U$ promotes antiferromagnetic (AFM) 
spin-density-wave (SDW) correlations \cite{Hirsch1985,White1989}, a $V$ term favors checkerboard charge-density-wave (CDW) order by penalizing adjacent charge fluctuations \cite{Solyom1979,Bari1971,Hirsch1984,Hirsch1985,Nakamura1999}. 
A central question is how this CDW tendency competes with the stripe manifold of the doped Hubbard model, and whether it can stabilize qualitatively new intertwined spin-charge order away from half filling.
In one dimension, 
this competition leads to a
surprisingly rich phase diagram, including CDW, SDW, and non-trivial bond-ordered-wave phases \cite{Nakamura2000}
separated by continuous or first-order transitions \cite{Sengupta2002}. These results highlight that even short-range
nonlocal interactions can substantially reorganize collective behaviors.

In two dimensions, the extended Hubbard model remains far less understood. Exact diagonalization on 
small clusters shows that sufficiently strong $V$ ($ \gtrsim U/4$) stabilizes CDW order at half-filling \cite{Ohta1994}, and various approximate methods have mapped aspects of the SDW–CDW competition\cite{Aichhorn2004,Davoudi2006,Paki2019}. 
While there have been some explorations in
doped systems\cite{Huang2013,Reymbaut2016, Jiang2018,Chen2023},
they have been mostly restricted to %in 
the context of superconducting instabilities, using 
approximate treatments. 
A number of key questions are 
unresolved, among them the  
fate of the ubiquitous stripe phases with extended 
interactions. 
How does a finite $V$ impact the robustness and possible evolution of the spin and charge order 
away from half filling? What is the interplay with longer-range hopping, e.g.,  $t^\prime$?
This regime is computationally very difficult. 
Even with 
only a $U$ term, determining the ground state in the Hubbard model required both high accuracy and access to large lattice sizes, as the underlying stripe orders are long-wavelength collective modes which
separated from other competing phases by very small energy scales. Adding a new competing scale
$V$ significantly increases the computational challenge. For example, in auxiliary-field quantum Monte Carlo (AFQMC) methods \cite{Zhang1995,Zhang1997}, the presence
of $V$ increases the number of auxiliary-fields and worsens the sign problem. In density matrix renormalization group (DMRG) \cite{White1992,White1993,Schollwock2005} or tensor network methods, extending the range of interaction increases the entanglement. 

 \begin{figure*}[t]
 \includegraphics[width=2.0\columnwidth]{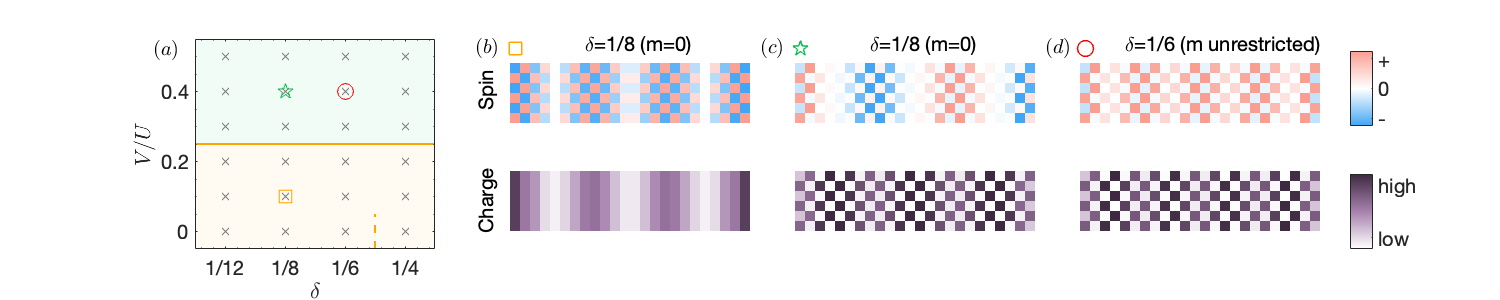}
\caption{\textbf{Ground-state phases of %Phase diagram of 
the doped extended-Hubbard model at $U/t=8$.
} 
(a) Phase diagram versus nearest-neighbor repulsion $V/U$ and hole doping $\delta$ up to $\delta\sim 1/4$. 
Crosses indicate parameter points where AFQMC calculations were performed. 
Representative spin and charge configurations are shown in (b)–(d). 
For $V/U \lesssim 0.25$, the ground state exhibits conventional AFM  
spin stripes (b), similar to the $t$--$U$ Hubbard model. 
For $V/U \gtrsim 0.25$, the system develops a modulated ferrimagnetic order intertwined with a checkerboard CDW. 
In the spin-balanced sector ($N_\uparrow=N_\downarrow$), positive--near-zero and negative--near-zero domains alternate in space (c); when spin polarization is unconstrained, the system selects a uniform positive--near-zero (or negative--near-zero) pattern with finite magnetization (d). 
The dashed orange curve in (a) indicates the critical doping at $V=0$ below which stripe order vanishes in the thermodynamic limit \cite{Xu2022}.
}
 \label{fig:phase_diagram}
 \end{figure*}
In this work, we address these questions, 
using two complementary computational methods: constrained-path (CP) AFQMC and DMRG.
The combination of these two high-accuracy many-body methods,
as we further discuss below, creates a synergy that 
allows us to reach better predictive power and higher confidence 
in determining the ground-state phases.
We find that a modest repulsion 
$V/U \gtrsim 0.25$ can qualitatively reorganize the stripe physics of the Hubbard model, 
stabilizing a modulated ferrimagnetic order intertwined with a checkerboard CDW. 
When overall spin is balanced, the order is 
modulated polarized spin stripes,
while in the spin-unrestricted case the system selects a ferrimagnetic realization with finite magnetization. 
The same ordering tendency persists in the presence of next-nearest-neighbor hopping $t'$, which tunes the modulation wavelength.
These findings are important to understanding cuprate materials --- and emergent phenomena in general. They also provide valuable guidance
for quantum simulation platforms. The simplicity of the Hamiltonian and the robustness of the phases we uncover make 
this system an excellent candidate for testing and realization with programmable quantum simulators.

%%%%%%%%%%%%%%%%%%%%%%%%%%%%%%%%%%%%%%%%%%%%%%%%%%%%%%%%%%%%%%%%%%%%%%%%%%%%%%%%%%%%%%%%%
%%%%%%%%%%%%%%%%%%%%%%%%%%%%%%%%%%%%%%%%%%%%%%%%%%%%%%%%%%%%%%%%%%%%%%%%%%%%%%%%%%%%%%%%%

\vskip0.10in
\noindent
\textit{Hamiltonian and Methodology.} 
The Extended Hubbard model with nearest neighbor interaction is defined by the Hamiltonian
\begin{equation}\label{eq:Hubbard_N1}
H = -t \sum_{\langle i,j \rangle, \sigma} \left( \hat{c}_{i \sigma}^\dagger \hat{c}_{j \sigma}^{\phantom{\dagger}} + \mathrm{h.c.} \right) + U \sum_{i} \hat{n}_{i \uparrow} \hat{n}_{i \downarrow} + V \sum_{\langle i,j \rangle}  \hat{n}_{i} \hat{n}_{j}\,,
\end{equation} 
where ${\langle i,j\rangle}$ denotes nearest-neighbors, $\hat{c}_{i \sigma}^\dagger$ ($\hat{c}_{i \sigma}^{\phantom{\dagger}} $) creates (annihilates) an electron with spin $\sigma$ on site $i$,
$\hat{n}_{i \sigma}^{\phantom{\dagger}} = \hat{c}_{i \sigma}^\dagger \hat{c}_{i \sigma}^{\phantom{\dagger}}$ is the number operator for spin $\sigma$. % on site $i$. 
The first term describes electron hopping and $t$ serves as the energy unit. The onsite and nearest-neighbor interaction strength are $U$ and $V$ respectively.
$N=L_x\times L_y$ is the number of lattice sites; typically we use open boundary condition for the long direction ($x$) and periodic boundary condition (PBC) for the short direction ($y$), although we have performed spot checks with PBC in both directions to ensure our overall results are not affected.  
To visualize stripe patterns we apply a weak boundary pinning field ($v_p=0.1$), which selects among nearly degenerate symmetry-related configurations without affecting bulk observables (see SM).
To characterize the magnetic orders, 
we measure the spin $\langle \hat{S}_i^z \rangle =\frac{1}{2}\langle \hat{n}_{i\uparrow} - \hat{n}_{i\downarrow}\rangle $ and charge densities $\langle \hat{n}_i \rangle =\langle \hat{n}_{i\uparrow} + \hat{n}_{i\downarrow}\rangle $  in real space. The total density of the system is %denoted as 
$ n  = 1/N \sum_{i} \langle \hat{n}_{i\uparrow} + \hat{n}_{i\downarrow} \rangle$, hence the doping is $\delta= 1-n$ or $n-1$ for hole/electron-doped %or electron-doped 
system, respectively. 
We also measure the spin and charge structure factor, $S=\frac{1}{ N}\sum_{i} \langle \hat{S}_i^{z}\rangle e^{ik\cdot r_i}$,and $C=\frac{1}{ N}\sum_{i} \langle \hat{n}_i -n\rangle e^{ik\cdot r_i}$ in momentum space via a Fourier Transform of spin and charge densities. 

\begin{figure}[t]
 \includegraphics[width=1\columnwidth]{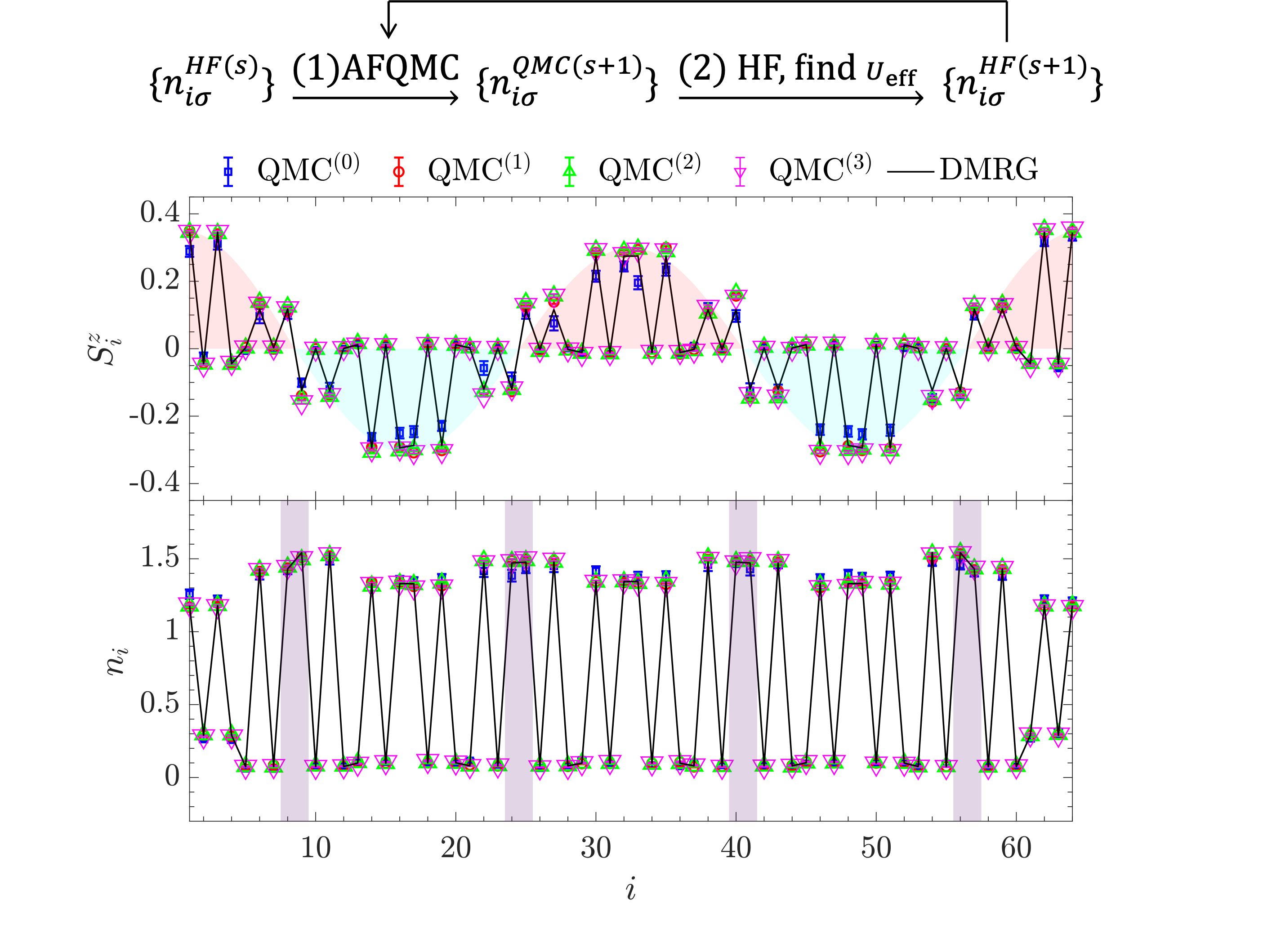}
\caption{\textbf{%(a) 
Schematic illustration of the self-consistent CP 
 AFQMC approach, and benchmark against DMRG.}
The top panel is a diagram of the self-consistent 
AFQMC procedure, in which  an effective HF trial wave function is used as the constraint. The middle and bottom panels show 
the computed spin and charge densities,
with four colors for the first four iterations of AFQMC and the solid line 
for DMRG. 
The calculation is 
at $U=8, V/U=0.4$, and $\delta=1/4$,
on a $16 \times 4$ rectangular lattice.
In the middle panel, the convention for site label $i$
(horizontal axis)
is depicted in the inset, and the 
pink and blue shaded regions highlight domains of alternating spin polarization. 
In the bottom panel, the purple shading indicates the nodes of the spin modulation from the middle panel; 
enhanced charge density 
is observed at these locations. 
Sites are indexed by $(i_x,i_y)\rightarrow
i_y+L_y\,(i_x-1)$.
}
 \label{fig:method_self_consistency}
 \end{figure}

We investigate the ground state properties 
using a combination of the CP AFQMC and DMRG
approaches. 
Both of these methods have been extensively 
benchmarked~\cite{LeBlanc2015,Zheng2017,Qin2020,Xu2023}, and their combined use has been especially fruitful in the context of Hubbard models~\cite{LeBlanc2015,Zheng2017,Qin2020,Xu2023}.
DMRG is a variational technique based on matrix product states and is particularly accurate for systems with low entanglement, such as quasi-one-dimensional systems with stronger interactions. 
In contrast, the CP AFQMC method is size-extensive, in that it can treat 
different boundary conditions and 2D (or higher) systems, with systematic errors (e.g., in $\langle H\rangle/N$) which are not sensitive 
to system size.
It relies on the CP approximation to control the sign problem, and 
its accuracy 
can be systematically improved by incorporating self-consistency procedures~\cite{Qin2016, Feng2023}, as discussed below. 
In this work, we employ these two complementary methods, using DMRG to benchmark CP AFQMC in smaller cylindrical systems, and then using % the latter 
AFQMC to access larger two-dimensional clusters and assess the stability of the ordering patterns.
In addition to the two types of many-body calculations, 
we also perform unrestricted and generalized Hartree Fock (UHF and GHF) calculations,  
details of which %the UHF 
are given in SM.
This serves several purposes: to suggest candidate 
orders in the system; to provide a reference for understanding correlation effects; to help gauge system size dependency; and to produce trial wave functions for AFQMC in our self-consistent CP scheme, 
as illustrated in the top panel of Fig.~\ref{fig:method_self_consistency}. 

In the self-consistent CP calculation, we couple the AFQMC calculation, which always uses the original many-body Hamiltonian $H$ in Eq.~(\ref{eq:Hubbard_N1}),
to a UHF or GHF
calculation with an {\it effective\/} Hartree-Fock (HF) Hamiltonian.
A HF wave function with $U_{\rm eff}$ and $V_{\rm eff}$ is used as the initial trial wave function at step $s=0$. CP-AFQMC then generates a set of output spin densities $\left \{n_{i\sigma}^{\rm QMC (s+1)} \right \}$, which are used as input for the next HF calculation. The parameters $U_{\rm eff}$ and $V_{\rm eff}$ are tuned such that the resulting HF 
spin densities$\left \{n_{i\sigma}^{\rm HF (s+1)} \right \}$ closely match   the input ones $\left \{n_{i\sigma}^{\rm QMC (s+1)} \right \}$. 
 This updated GHF 
 solution  $\left \{n_{i\sigma}^{\rm HF (s+1)} \right \}$  is then used as the trial wave function for the next iteration of the QMC calculation. The process is iterated until 
 convergence.
Convergence to the same final state from different initial trial wave functions derived from different states provides 
a strong indicator of the robustness of the 
 final solution(see SM).
%%%%%%%%%%%%%%%%%%%%%%%%%%%%%%%%%%%%%%%%%%%%%%%%%%%%%%%%%%%%%%%%%%%%%%%%%%%%%%%%%%%%%%%%%%%%%%%%%%%%%%%%%%%%%%

\vskip0.10in
\noindent
\textit{Results.} 
Figure~\ref{fig:phase_diagram} summarizes the ground-state behavior as a function of $V/U$ and doping $\delta$ at $U/t=8$. 
We study cylindrical lattices up to $L_x=36$ and $L_y=12$; finite-size effects on the reported ordering patterns are limited (see SM). 
A qualitative reorganization occurs near a threshold $V_c/U \sim 0.25$. 
Below it, we obtain the familiar 
AFM spin-stripe ground state of the doped Hubbard model. 
Above it, 
we find a modulated ferrimagnetic order intertwined with a checkerboard CDW.

The realization of this intertwined order depends on the spin constraint. 
In the spin-balanced sector ($N_\uparrow=N_\downarrow$), domains with opposite net polarization alternate in space, producing an antiferromagnetic arrangement of positive--near-zero and negative--near-zero stripe domains. 
When spin polarization is unconstrained, the system selects a ferrimagnetic realization with finite magnetization. 
These variants are close in energy and can be tuned experimentally, e.g., by controlling spin imbalance in cold-atom platforms~\cite{Zwierlein2006,Partridge2006a,Shin2006,Schunck2007a,Shin2008,Mitra2016,Schirotzek2009,Liao2010,Massignan2025,Feng2025, Hartke2025}. 

Representative spin and charge density patterns for different dopings 
are illustrated 
in Fig.~\ref{fig:phase_diagram}(b)-(d).
Below a threshold of $V/U \sim 0.25$, %For $V/U < 0.25$, 
the system exhibits a spin stripe phase with wavelength $\lambda = 2/\delta$, accompanied by a dip in charge density (i.e., a peak in hole density)
at the interfaces between different AFM 
domains, as depicted in Fig.~\ref{fig:phase_diagram}(b).
This is the same state as seen in the usual Hubbard model when $V=0$. 
In contrast, 
above the threshold, 
a novel type of magnetic order appears.
As illustrated in Fig.~\ref{fig:phase_diagram}(c) (and also Fig.~\ref{fig:method_self_consistency}), this state is
characterized by 
a modulated ferrimagnetic stripe texture intertwined with a checkerboard CDW.
For fixed $N_{\uparrow}=N_{\downarrow}$,
the global pattern displays an alternating sequence of domains with positive
and negative 
magnetizations. We also find another nearly degenerate ground state in which  domains of the same sign 
cluster together and are separated from the 
opposite domains. 
When the total magnetization is allowed to vary, 
a ferrimagnetic phase is nearly degenerate in energy for $\delta \gtrsim 1/6$ and become slightly favored $\delta = 1/4$ (see SM for further details on the energetics and near degeneracies).
In this state, shown in Fig.~\ref{fig:phase_diagram}(d),
$n=\delta/2$ spins flip from down to up, yielding a spin-up density of $1/2$, the maximum allowed in a staggered pattern without incurring penalties due to nearest-neighbor repulsion $V$.

Before characterizing these phases more quantitatively, we illustrate our computational approaches, with an outline
of the self-consistent CP AFQMC and a comparison between it and DMRG, in Fig.~\ref{fig:method_self_consistency}. 
The upper panel sketches the self-consistency procedure in the CP-AFQMC. In the lower panels, we plot 
the computed spin ($S_i^z$) and charge ($n_i$) densities along site index $i$ on a $16 \times 4$ lattice, 
from both DMRG and the first several iterations of AFQMC.
Convergence is reached within a few iterations, and  excellent agreement is seen 
between the converged result and DMRG. 
This benchmark demonstrates that the intertwined order is not an artifact of the constraint or trial wave function, but a robust correlated ground-state feature.
The ground state for this particular system is similar to
that depicted in Fig.~\ref{fig:phase_diagram}(c).
Alternating polarized spin stripes emerge: within each domain, spin densities alternate between positive and almost zero (pink regions) or negative and almost zero (blue regions).
A robust CDW pattern develops across the entire lattice. Charge density oscillations are particularly enhanced at domain boundaries (indicated by purple shading). 

 \begin{figure}[t]
 \includegraphics[width=1\columnwidth]{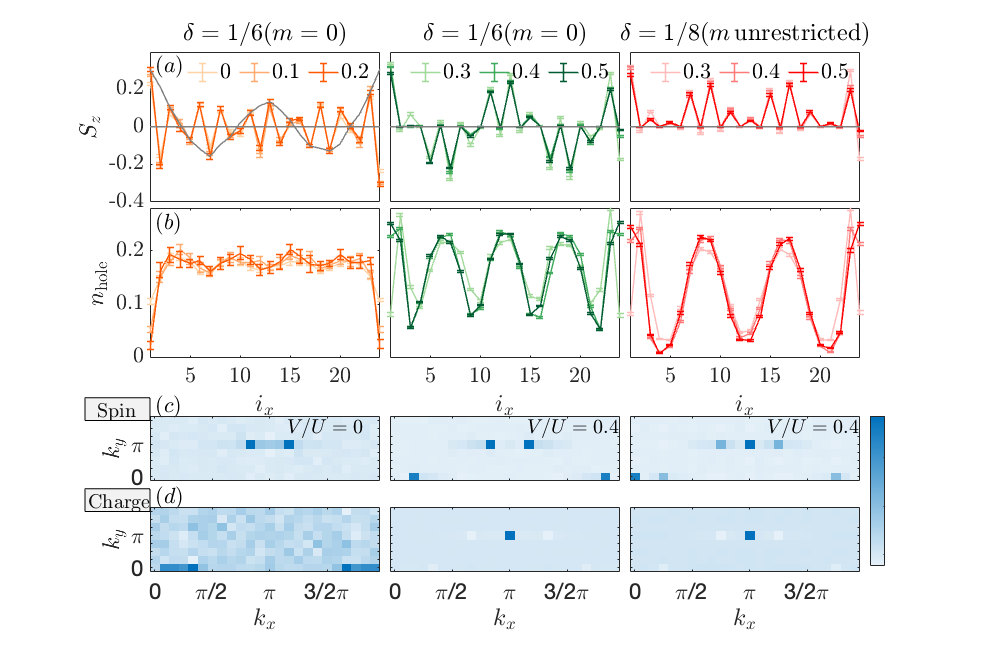}
\caption{\textbf{Characterization of the magnetic and charge properties in the different phases.}
The left, middle, and right panels correspond
to the three phases shown in Fig.~\ref{fig:phase_diagram}b, Fig.~\ref{fig:phase_diagram}c, and Fig.~\ref{fig:phase_diagram}d, respectively, following the same color schemes.
The top two rows show spin and hole density line cuts, with the shades of color in the different 
curves in each indicating different values of $V/U$.
In the botton panel, the spin
and charge structure factors are plotted in momentum-space for the three phases, each at a selected value of $V/U$ (shown in the legends in 
the spin plots). 
All simulations are performed on a $24 \times 8$ lattice with $U=8$.
}
 \label{fig:k_space}
 \end{figure}

To further quantify 
these phases, Fig.~\ref{fig:k_space} presents the spin and charge properties as a function of nearest neighbor interaction strength $V/U$. Spin densities linecuts and hole densities (average over all rows) are plotted along the site index $i_x$ in Fig.~\ref{fig:k_space}(a)(b). When $V/U \lesssim 0.25$, the system exhibits spin stripes (left) with a wavelength $\lambda=2/\delta$, with hole accumulated at the node of the spin stripe. Consequently, the accompanying charge modulation has a wavelength $\lambda=1/\delta$. As $V/U$ increases, oscillations in spin densities and charge modulation remain nearly unchanged except near the boundaries.
When 
$V/U$ exceeds  
$\sim 0.25$, the new ferrimagnetic state is seen.
There are several nearly-degenerate forms of this state which are energetically difficult to resolve, as we have discussed.  
When the total magnetization $m$ is constrained to zero, the most typical state seen, especially at smaller values of $V$ and low doping, is the one in which 
the spin and charge modulations maintain the same wavelengths as in the spin stripe phase, but the holes now accumulate at the maxima of the spin amplitude instead of the nodes. A state in which the modulation wavelength is 
larger, i.e., with a smaller 
number of ferrimagnetic domains,
is nearly degenerate.(see SM)
When the total magnetization is allowed to vary, 
a global ferrimagnetic order emerges, with coexisting CDW. 
The right panels show 
spin and charge patterns with wavelength of $\lambda=1/\delta$. 
Note that the staggered spin density does not cross the zero line, indicating the absence of a $\pi$ phase shift between neighboring domains. 
In all these cases, the change in 
spin and charge amplitudes with 
$V/U$ remains modest, as seen in the middle and right panels in the top two rows.

Panels (c) and (d) display spin and charge structure factors, i.e. the Fourier transforms of the corresponding real-space densities for the three phases. At $V/U=0$, the system resides in the spin stripe phase with wavelength $\lambda = 2/\delta$, giving rise to a peak at $((1-\delta)\pi, \pi) = (5\pi/6, \pi)$. In the charge channel, holes accumulate at AFM domain boundaries, leading to a modulation with wavelength $\lambda = 1/\delta$, and a corresponding peak at $(2\delta\pi, 0)$ in the momentum space. When $V/U=0.4$ and $N_{\uparrow}=N_{\downarrow}$ are fixed, the peak at $((1-\delta)\pi, \pi)$ persists, reflecting the alternating spin texture (positive/negative and zero) within domains. A new peak emerges at $(\delta\pi, 0) = (\pi/6, 0)$, signaling the periodic alternation between domains. Coexisting CDW order manifests as a dominant peak at $(\pi, \pi)$. Meanwhile, in real space, holes accumulate at the maxima of spin densities, which appears as a dip at $((1-2\delta)\pi,\pi)$ in momentum space. When the spin degrees of freedom are released, a ferrimagnetic order emerges together with CDW order. Thus a peak at $(0, 0)$ develops in the spin channel, indicating a finite total magnetization, while the peak at $(\pi, \pi)$ corresponds to the overall staggered spin pattern without any $\pi$ phase-shifted domains as seen in the other two phases. Additional peaks at $((1-\delta)\pi, \pi)$ and $(\delta \pi,0)$ reflect the periodicity $\lambda=1/\delta$ along the horizontal direction. In the charge channel, the peaks and dips appear at the same positions as in the middle panel. 

\begin{figure*}[t]
 \includegraphics[width=2\columnwidth]{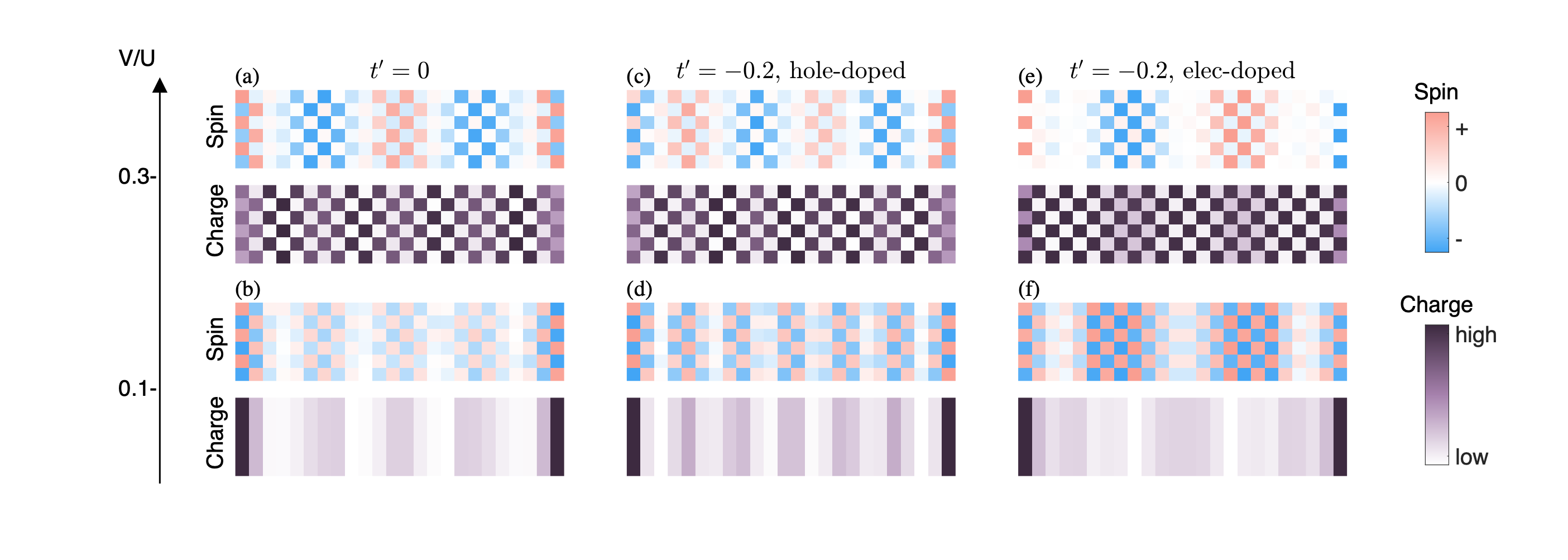}
\caption{\textbf{Effect of next-nearest-neighbor hopping $t^\prime$ on spin and charge order.}
Spin and charge density profiles are shown for 
two values of $V/U$  ($0.1$ in the lower half and $0.3$ in the upper half), for three values of 
next-near-neighbor hopping $t^\prime/t$,
representing $t^\prime = 0$ as studied above 
and hole- and electron-doped cases, respectively.
Partially filled stripes appear in the hole-doped case, while overfilled stripes are seen under electron doping when $t^\prime = -0.2$.
The transition from pure spin stripes to the coexistence of charge density waves (CDW) and alternating polarized spin stripes occurs near $V/U \sim 0.25$, independent of the sign of doping or the value of $t^\prime$.
All calculations are performed at doping $\delta = 1/6$ and $U/t=8$.
}
 \label{fig:t_prime_effects}
 \end{figure*}
Finally we explore the effects of next-nearest-neighbor hopping term $H_{t^{\prime}}=-t^{\prime} \sum_{\langle \langle i,j \rangle \rangle, \sigma} \left( \hat{c}_{i \sigma}^\dagger \hat{c}_{j \sigma}^{\phantom{\dagger}} + \mathrm{h.c.} \right)$ in Fig.~\ref{fig:t_prime_effects}. Panels (a) and (b) show spin and charge configurations at $t'=0$ for $V/U=0.1$ and $0.3$, serving as a reference. As discussed earlier, the former exhibits spin stripes with charge modulations, while the latter shows alternating polarized spin stripes and CDW patterns. When $t'$ is introduced, electron- and hole-doped regimes behave differently due to the absence of particle-hole symmetry (panels c–f). In the hole-doped case, partially filled spin stripes show up for $V/U \lesssim 0.25$, consistent with the pure Hubbard model with $V=0$ at finite $t'=-0.2$ \cite{Xu2023}. Notably, $t'$ controls the number of stripes without altering the fundamental spin order. When $V/U \gtrsim 0.25$, alternating polarized spin stripes coexist with CDW patterns, with a shorter wavelength compared with the $t'=0$ case. The transition from spin stripes to alternating magnetic order still occurs near $V_c/U \sim 0.25$, mirroring the $t'=0$ case. On the electron-doped side, the number of stripes decreases (i.e., the wavelength increases) compared to the $t'=0$ case, resulting in overfilled spin-stripe states. The longer wavelength makes the stripe order more difficult to observe, and long-range antiferromagnetic (AFM) order is expected to dominate in experiments, especially at low doping \cite{Xu2023}. For $V/U \lesssim 0.25$, excess electrons accumulate near the stripe nodes. A similar transition occurs around $V_c/U \sim 0.25$, above which polarized spin stripes coexist with CDW order.

\textit{Conclusions.} 
In this work, we have investigated the effects of nearest-neighbor interactions in the Hubbard model using two complementary high-precision many-body approaches.
We uncover a new phase, a modulated ferrimagnetic order intertwined with a
checkerboard CDW, when the strength 
$V/U \gtrsim 0.25$.
This phase can manifest itself 
in several forms depending on 
the constraint on global spin polarization, details of the parameters, and even finite-size effects.
The next-nearest-neighbor hopping primarily tunes the stripe periodicity, while leaving the transition between conventional spin stripes and the new 
phase 
essentially unchanged.

Our results uncover novel magnetic orders in the extended Hubbard model and highlight the crucial role of longer-range interactions in shaping the collective behavior of strongly correlated systems. The emergence of polarized and ferrimagnetic stripe phases provides a new platform for studying the interplay between spin, charge, and magnetization in correlated materials, and raises important questions about their potential impact on superconducting instabilities. Importantly, our findings are directly relevant to ongoing experimental efforts in quantum simulation\cite{Xu2025}. Modern cold-atom platforms, including dipolar gases and Rydberg-dressed systems\cite{Guardado-Sanchez2021,Langen2024,Cornish2024}, as well as weakly screened moiré heterostructures\cite{XuQiaoling2025}, offer realistic routes to realize the Hubbard model with tunable nearest-neighbor interactions on cylindrical geometries. Our work thus provides concrete guidance for the experimental exploration of interaction-driven stripe phases and correlated orders beyond the conventional Hubbard paradigm.

\vskip0.20in \noindent
\textit{Acknowledgements-}  
We thank Ryan Levy, Yiqi Yang, and Kyle Eskridge for providing useful benchmark and helpful discussion. We thank the Flatiron Institute Scientific Computing Center for computational resources. The Flatiron Institute is a division of the Simons Foundation.
\bibliography{Extended_Hubbard_with_V}

\begin{thebibliography}{50}%
\makeatletter
\providecommand \@ifxundefined [1]{%
 \@ifx{#1\undefined}
}%
\providecommand \@ifnum [1]{%
 \ifnum #1\expandafter \@firstoftwo
 \else \expandafter \@secondoftwo
 \fi
}%
\providecommand \@ifx [1]{%
 \ifx #1\expandafter \@firstoftwo
 \else \expandafter \@secondoftwo
 \fi
}%
\providecommand \natexlab [1]{#1}%
\providecommand \enquote  [1]{``#1''}%
\providecommand \bibnamefont  [1]{#1}%
\providecommand \bibfnamefont [1]{#1}%
\providecommand \citenamefont [1]{#1}%
\providecommand \href@noop [0]{\@secondoftwo}%
\providecommand \href [0]{\begingroup \@sanitize@url \@href}%
\providecommand \@href[1]{\@@startlink{#1}\@@href}%
\providecommand \@@href[1]{\endgroup#1\@@endlink}%
\providecommand \@sanitize@url [0]{\catcode `\\12\catcode `\$12\catcode
  `\&12\catcode `\#12\catcode `\^12\catcode `\_12\catcode `\%12\relax}%
\providecommand \@@startlink[1]{}%
\providecommand \@@endlink[0]{}%
\providecommand \url  [0]{\begingroup\@sanitize@url \@url }%
\providecommand \@url [1]{\endgroup\@href {#1}{\urlprefix }}%
\providecommand \urlprefix  [0]{URL }%
\providecommand \Eprint [0]{\href }%
\providecommand \doibase [0]{https://doi.org/}%
\providecommand \selectlanguage [0]{\@gobble}%
\providecommand \bibinfo  [0]{\@secondoftwo}%
\providecommand \bibfield  [0]{\@secondoftwo}%
\providecommand \translation [1]{[#1]}%
\providecommand \BibitemOpen [0]{}%
\providecommand \bibitemStop [0]{}%
\providecommand \bibitemNoStop [0]{.\EOS\space}%
\providecommand \EOS [0]{\spacefactor3000\relax}%
\providecommand \BibitemShut  [1]{\csname bibitem#1\endcsname}%
\let\auto@bib@innerbib\@empty
%</preamble>
\bibitem [{\citenamefont {Hubbard}\ and\ \citenamefont
  {Flowers}(1963)}]{Hubbard1963}%
  \BibitemOpen
  \bibfield  {author} {\bibinfo {author} {\bibfnamefont {J.}~\bibnamefont
  {Hubbard}}\ and\ \bibinfo {author} {\bibfnamefont {B.~H.}\ \bibnamefont
  {Flowers}},\ }\bibfield  {title} {\bibinfo {title} {{Electron correlations in
  narrow energy bands}},\ }\href {https://doi.org/10.1098/rspa.1963.0204}
  {\bibfield  {journal} {\bibinfo  {journal} {Proceedings of the Royal Society
  of London. Series A. Mathematical and Physical Sciences}\ }\textbf {\bibinfo
  {volume} {276}},\ \bibinfo {pages} {238} (\bibinfo {year}
  {1963})}\BibitemShut {NoStop}%
\bibitem [{\citenamefont {Kivelson}\ \emph {et~al.}(2003)\citenamefont
  {Kivelson}, \citenamefont {Bindloss}, \citenamefont {Fradkin}, \citenamefont
  {Oganesyan}, \citenamefont {Tranquada}, \citenamefont {Kapitulnik},\ and\
  \citenamefont {Howald}}]{Kivelson2003}%
  \BibitemOpen
  \bibfield  {author} {\bibinfo {author} {\bibfnamefont {S.~A.}\ \bibnamefont
  {Kivelson}}, \bibinfo {author} {\bibfnamefont {I.~P.}\ \bibnamefont
  {Bindloss}}, \bibinfo {author} {\bibfnamefont {E.}~\bibnamefont {Fradkin}},
  \bibinfo {author} {\bibfnamefont {V.}~\bibnamefont {Oganesyan}}, \bibinfo
  {author} {\bibfnamefont {J.~M.}\ \bibnamefont {Tranquada}}, \bibinfo {author}
  {\bibfnamefont {A.}~\bibnamefont {Kapitulnik}},\ and\ \bibinfo {author}
  {\bibfnamefont {C.}~\bibnamefont {Howald}},\ }\bibfield  {title} {\bibinfo
  {title} {How to detect fluctuating stripes in the high-temperature
  superconductors},\ }\href {https://doi.org/10.1103/RevModPhys.75.1201}
  {\bibfield  {journal} {\bibinfo  {journal} {Rev. Mod. Phys.}\ }\textbf
  {\bibinfo {volume} {75}},\ \bibinfo {pages} {1201} (\bibinfo {year}
  {2003})}\BibitemShut {NoStop}%
\bibitem [{\citenamefont {Proust}\ and\ \citenamefont
  {Taillefer}(2019)}]{Proust2019}%
  \BibitemOpen
  \bibfield  {author} {\bibinfo {author} {\bibfnamefont {C.}~\bibnamefont
  {Proust}}\ and\ \bibinfo {author} {\bibfnamefont {L.}~\bibnamefont
  {Taillefer}},\ }\bibfield  {title} {\bibinfo {title} {The remarkable
  underlying ground states of cuprate superconductors},\ }\href
  {https://doi.org/https://doi.org/10.1146/annurev-conmatphys-031218-013210}
  {\bibfield  {journal} {\bibinfo  {journal} {Annual Review of Condensed Matter
  Physics}\ }\textbf {\bibinfo {volume} {10}},\ \bibinfo {pages} {409}
  (\bibinfo {year} {2019})}\BibitemShut {NoStop}%
\bibitem [{\citenamefont {Arovas}\ \emph {et~al.}(2022)\citenamefont {Arovas},
  \citenamefont {Berg}, \citenamefont {Kivelson},\ and\ \citenamefont
  {Raghu}}]{Arovas2022}%
  \BibitemOpen
  \bibfield  {author} {\bibinfo {author} {\bibfnamefont {D.~P.}\ \bibnamefont
  {Arovas}}, \bibinfo {author} {\bibfnamefont {E.}~\bibnamefont {Berg}},
  \bibinfo {author} {\bibfnamefont {S.~A.}\ \bibnamefont {Kivelson}},\ and\
  \bibinfo {author} {\bibfnamefont {S.}~\bibnamefont {Raghu}},\ }\bibfield
  {title} {\bibinfo {title} {The {H}ubbard model},\ }\href
  {https://doi.org/https://doi.org/10.1146/annurev-conmatphys-031620-102024}
  {\bibfield  {journal} {\bibinfo  {journal} {Annual Review of Condensed Matter
  Physics}\ }\textbf {\bibinfo {volume} {13}},\ \bibinfo {pages} {239}
  (\bibinfo {year} {2022})}\BibitemShut {NoStop}%
\bibitem [{\citenamefont {Qin}\ \emph {et~al.}(2020)\citenamefont {Qin},
  \citenamefont {Chung}, \citenamefont {Shi}, \citenamefont {Vitali},
  \citenamefont {Hubig}, \citenamefont {Schollw\"ock}, \citenamefont {White},\
  and\ \citenamefont {Zhang}}]{Qin2020}%
  \BibitemOpen
  \bibfield  {author} {\bibinfo {author} {\bibfnamefont {M.}~\bibnamefont
  {Qin}}, \bibinfo {author} {\bibfnamefont {C.-M.}\ \bibnamefont {Chung}},
  \bibinfo {author} {\bibfnamefont {H.}~\bibnamefont {Shi}}, \bibinfo {author}
  {\bibfnamefont {E.}~\bibnamefont {Vitali}}, \bibinfo {author} {\bibfnamefont
  {C.}~\bibnamefont {Hubig}}, \bibinfo {author} {\bibfnamefont
  {U.}~\bibnamefont {Schollw\"ock}}, \bibinfo {author} {\bibfnamefont {S.~R.}\
  \bibnamefont {White}},\ and\ \bibinfo {author} {\bibfnamefont
  {S.}~\bibnamefont {Zhang}} (\bibinfo {collaboration} {Simons Collaboration on
  the Many-Electron Problem}),\ }\bibfield  {title} {\bibinfo {title} {Absence
  of superconductivity in the pure two-dimensional {H}ubbard model},\ }\href
  {https://doi.org/10.1103/PhysRevX.10.031016} {\bibfield  {journal} {\bibinfo
  {journal} {Phys. Rev. X}\ }\textbf {\bibinfo {volume} {10}},\ \bibinfo
  {pages} {031016} (\bibinfo {year} {2020})}\BibitemShut {NoStop}%
\bibitem [{\citenamefont {Wietek}\ \emph {et~al.}(2021)\citenamefont {Wietek},
  \citenamefont {He}, \citenamefont {White}, \citenamefont {Georges},\ and\
  \citenamefont {Stoudenmire}}]{Wietek2021}%
  \BibitemOpen
  \bibfield  {author} {\bibinfo {author} {\bibfnamefont {A.}~\bibnamefont
  {Wietek}}, \bibinfo {author} {\bibfnamefont {Y.-Y.}\ \bibnamefont {He}},
  \bibinfo {author} {\bibfnamefont {S.~R.}\ \bibnamefont {White}}, \bibinfo
  {author} {\bibfnamefont {A.}~\bibnamefont {Georges}},\ and\ \bibinfo {author}
  {\bibfnamefont {E.~M.}\ \bibnamefont {Stoudenmire}},\ }\bibfield  {title}
  {\bibinfo {title} {Stripes, antiferromagnetism, and the pseudogap in the
  doped {H}ubbard model at finite temperature},\ }\href
  {https://doi.org/10.1103/PhysRevX.11.031007} {\bibfield  {journal} {\bibinfo
  {journal} {Phys. Rev. X}\ }\textbf {\bibinfo {volume} {11}},\ \bibinfo
  {pages} {031007} (\bibinfo {year} {2021})}\BibitemShut {NoStop}%
\bibitem [{\citenamefont {Xu}\ \emph {et~al.}(2024)\citenamefont {Xu},
  \citenamefont {Chung}, \citenamefont {Qin}, \citenamefont {Schollwöck},
  \citenamefont {White},\ and\ \citenamefont {Zhang}}]{Xu2023}%
  \BibitemOpen
  \bibfield  {author} {\bibinfo {author} {\bibfnamefont {H.}~\bibnamefont
  {Xu}}, \bibinfo {author} {\bibfnamefont {C.-M.}\ \bibnamefont {Chung}},
  \bibinfo {author} {\bibfnamefont {M.}~\bibnamefont {Qin}}, \bibinfo {author}
  {\bibfnamefont {U.}~\bibnamefont {Schollwöck}}, \bibinfo {author}
  {\bibfnamefont {S.~R.}\ \bibnamefont {White}},\ and\ \bibinfo {author}
  {\bibfnamefont {S.}~\bibnamefont {Zhang}},\ }\bibfield  {title} {\bibinfo
  {title} {Coexistence of superconductivity with partially filled stripes in
  the {H}ubbard model},\ }\href {https://doi.org/10.1126/science.adh7691}
  {\bibfield  {journal} {\bibinfo  {journal} {Science}\ }\textbf {\bibinfo
  {volume} {384}},\ \bibinfo {pages} {eadh7691} (\bibinfo {year}
  {2024})}\BibitemShut {NoStop}%
\bibitem [{\citenamefont {Šimkovic}\ \emph {et~al.}(2024)\citenamefont
  {Šimkovic}, \citenamefont {Rossi}, \citenamefont {Georges},\ and\
  \citenamefont {Ferrero}}]{Simkovic2024}%
  \BibitemOpen
  \bibfield  {author} {\bibinfo {author} {\bibfnamefont {F.}~\bibnamefont
  {Šimkovic}}, \bibinfo {author} {\bibfnamefont {R.}~\bibnamefont {Rossi}},
  \bibinfo {author} {\bibfnamefont {A.}~\bibnamefont {Georges}},\ and\ \bibinfo
  {author} {\bibfnamefont {M.}~\bibnamefont {Ferrero}},\ }\bibfield  {title}
  {\bibinfo {title} {Origin and fate of the pseudogap in the doped {H}ubbard
  model},\ }\href {https://doi.org/10.1126/science.ade9194} {\bibfield
  {journal} {\bibinfo  {journal} {Science}\ }\textbf {\bibinfo {volume}
  {385}},\ \bibinfo {pages} {eade9194} (\bibinfo {year} {2024})}\BibitemShut
  {NoStop}%
\bibitem [{\citenamefont {Hirsch}(1985)}]{Hirsch1985}%
  \BibitemOpen
  \bibfield  {author} {\bibinfo {author} {\bibfnamefont {J.~E.}\ \bibnamefont
  {Hirsch}},\ }\bibfield  {title} {\bibinfo {title} {Phase diagram of the
  one-dimensional molecular-crystal model with {C}oulomb interactions:
  Half-filled-band sector},\ }\href {https://doi.org/10.1103/PhysRevB.31.6022}
  {\bibfield  {journal} {\bibinfo  {journal} {Phys. Rev. B}\ }\textbf {\bibinfo
  {volume} {31}},\ \bibinfo {pages} {6022} (\bibinfo {year}
  {1985})}\BibitemShut {NoStop}%
\bibitem [{\citenamefont {White}\ \emph {et~al.}(1989)\citenamefont {White},
  \citenamefont {Scalapino}, \citenamefont {Sugar}, \citenamefont {Loh},
  \citenamefont {Gubernatis},\ and\ \citenamefont {Scalettar}}]{White1989}%
  \BibitemOpen
  \bibfield  {author} {\bibinfo {author} {\bibfnamefont {S.~R.}\ \bibnamefont
  {White}}, \bibinfo {author} {\bibfnamefont {D.~J.}\ \bibnamefont
  {Scalapino}}, \bibinfo {author} {\bibfnamefont {R.~L.}\ \bibnamefont
  {Sugar}}, \bibinfo {author} {\bibfnamefont {E.~Y.}\ \bibnamefont {Loh}},
  \bibinfo {author} {\bibfnamefont {J.~E.}\ \bibnamefont {Gubernatis}},\ and\
  \bibinfo {author} {\bibfnamefont {R.~T.}\ \bibnamefont {Scalettar}},\
  }\bibfield  {title} {\bibinfo {title} {Numerical study of the two-dimensional
  {H}ubbard model},\ }\href {https://doi.org/10.1103/PhysRevB.40.506}
  {\bibfield  {journal} {\bibinfo  {journal} {Phys. Rev. B}\ }\textbf {\bibinfo
  {volume} {40}},\ \bibinfo {pages} {506} (\bibinfo {year} {1989})}\BibitemShut
  {NoStop}%
\bibitem [{\citenamefont {Sólyom}(1979)}]{Solyom1979}%
  \BibitemOpen
  \bibfield  {author} {\bibinfo {author} {\bibfnamefont {J.}~\bibnamefont
  {Sólyom}},\ }\bibfield  {title} {\bibinfo {title} {The {F}ermi gas model of
  one-dimensional conductors},\ }\href
  {https://doi.org/10.1080/00018737900101375} {\bibfield  {journal} {\bibinfo
  {journal} {Advances in Physics}\ }\textbf {\bibinfo {volume} {28}},\ \bibinfo
  {pages} {201} (\bibinfo {year} {1979})}\BibitemShut {NoStop}%
\bibitem [{\citenamefont {Bari}(1971)}]{Bari1971}%
  \BibitemOpen
  \bibfield  {author} {\bibinfo {author} {\bibfnamefont {R.~A.}\ \bibnamefont
  {Bari}},\ }\bibfield  {title} {\bibinfo {title} {Effects of short-range
  interactions on electron-charge ordering and lattice distortions in the
  localized state},\ }\href {https://doi.org/10.1103/PhysRevB.3.2662}
  {\bibfield  {journal} {\bibinfo  {journal} {Phys. Rev. B}\ }\textbf {\bibinfo
  {volume} {3}},\ \bibinfo {pages} {2662} (\bibinfo {year} {1971})}\BibitemShut
  {NoStop}%
\bibitem [{\citenamefont {Hirsch}(1984)}]{Hirsch1984}%
  \BibitemOpen
  \bibfield  {author} {\bibinfo {author} {\bibfnamefont {J.~E.}\ \bibnamefont
  {Hirsch}},\ }\bibfield  {title} {\bibinfo {title} {Charge-density-wave to
  spin-density-wave transition in the extended {H}ubbard model},\ }\href
  {https://doi.org/10.1103/PhysRevLett.53.2327} {\bibfield  {journal} {\bibinfo
   {journal} {Phys. Rev. Lett.}\ }\textbf {\bibinfo {volume} {53}},\ \bibinfo
  {pages} {2327} (\bibinfo {year} {1984})}\BibitemShut {NoStop}%
\bibitem [{\citenamefont {Nakamura}(1999)}]{Nakamura1999}%
  \BibitemOpen
  \bibfield  {author} {\bibinfo {author} {\bibfnamefont {M.}~\bibnamefont
  {Nakamura}},\ }\bibfield  {title} {\bibinfo {title} {Mechanism of {CDW-SDW}
  transition in one dimension},\ }\href {https://doi.org/10.1143/JPSJ.68.3123}
  {\bibfield  {journal} {\bibinfo  {journal} {Journal of the Physical Society
  of Japan}\ }\textbf {\bibinfo {volume} {68}},\ \bibinfo {pages} {3123}
  (\bibinfo {year} {1999})}\BibitemShut {NoStop}%
\bibitem [{\citenamefont {Nakamura}(2000)}]{Nakamura2000}%
  \BibitemOpen
  \bibfield  {author} {\bibinfo {author} {\bibfnamefont {M.}~\bibnamefont
  {Nakamura}},\ }\bibfield  {title} {\bibinfo {title} {Tricritical behavior in
  the extended {H}ubbard chains},\ }\href
  {https://doi.org/10.1103/PhysRevB.61.16377} {\bibfield  {journal} {\bibinfo
  {journal} {Phys. Rev. B}\ }\textbf {\bibinfo {volume} {61}},\ \bibinfo
  {pages} {16377} (\bibinfo {year} {2000})}\BibitemShut {NoStop}%
\bibitem [{\citenamefont {Sengupta}\ \emph {et~al.}(2002)\citenamefont
  {Sengupta}, \citenamefont {Sandvik},\ and\ \citenamefont
  {Campbell}}]{Sengupta2002}%
  \BibitemOpen
  \bibfield  {author} {\bibinfo {author} {\bibfnamefont {P.}~\bibnamefont
  {Sengupta}}, \bibinfo {author} {\bibfnamefont {A.~W.}\ \bibnamefont
  {Sandvik}},\ and\ \bibinfo {author} {\bibfnamefont {D.~K.}\ \bibnamefont
  {Campbell}},\ }\bibfield  {title} {\bibinfo {title} {Bond-order-wave phase
  and quantum phase transitions in the one-dimensional extended {H}ubbard
  model},\ }\href {https://doi.org/10.1103/PhysRevB.65.155113} {\bibfield
  {journal} {\bibinfo  {journal} {Phys. Rev. B}\ }\textbf {\bibinfo {volume}
  {65}},\ \bibinfo {pages} {155113} (\bibinfo {year} {2002})}\BibitemShut
  {NoStop}%
\bibitem [{\citenamefont {Ohta}\ \emph {et~al.}(1994)\citenamefont {Ohta},
  \citenamefont {Tsutsui}, \citenamefont {Koshibae},\ and\ \citenamefont
  {Maekawa}}]{Ohta1994}%
  \BibitemOpen
  \bibfield  {author} {\bibinfo {author} {\bibfnamefont {Y.}~\bibnamefont
  {Ohta}}, \bibinfo {author} {\bibfnamefont {K.}~\bibnamefont {Tsutsui}},
  \bibinfo {author} {\bibfnamefont {W.}~\bibnamefont {Koshibae}},\ and\
  \bibinfo {author} {\bibfnamefont {S.}~\bibnamefont {Maekawa}},\ }\bibfield
  {title} {\bibinfo {title} {Exact-diagonalization study of the {H}ubbard model
  with nearest-neighbor repulsion},\ }\href
  {https://doi.org/10.1103/PhysRevB.50.13594} {\bibfield  {journal} {\bibinfo
  {journal} {Phys. Rev. B}\ }\textbf {\bibinfo {volume} {50}},\ \bibinfo
  {pages} {13594} (\bibinfo {year} {1994})}\BibitemShut {NoStop}%
\bibitem [{\citenamefont {Aichhorn}\ \emph {et~al.}(2004)\citenamefont
  {Aichhorn}, \citenamefont {Evertz}, \citenamefont {von~der Linden},\ and\
  \citenamefont {Potthoff}}]{Aichhorn2004}%
  \BibitemOpen
  \bibfield  {author} {\bibinfo {author} {\bibfnamefont {M.}~\bibnamefont
  {Aichhorn}}, \bibinfo {author} {\bibfnamefont {H.~G.}\ \bibnamefont
  {Evertz}}, \bibinfo {author} {\bibfnamefont {W.}~\bibnamefont {von~der
  Linden}},\ and\ \bibinfo {author} {\bibfnamefont {M.}~\bibnamefont
  {Potthoff}},\ }\bibfield  {title} {\bibinfo {title} {Charge ordering in
  extended {H}ubbard models: Variational cluster approach},\ }\href
  {https://doi.org/10.1103/PhysRevB.70.235107} {\bibfield  {journal} {\bibinfo
  {journal} {Phys. Rev. B}\ }\textbf {\bibinfo {volume} {70}},\ \bibinfo
  {pages} {235107} (\bibinfo {year} {2004})}\BibitemShut {NoStop}%
\bibitem [{\citenamefont {Davoudi}\ and\ \citenamefont
  {Tremblay}(2006)}]{Davoudi2006}%
  \BibitemOpen
  \bibfield  {author} {\bibinfo {author} {\bibfnamefont {B.}~\bibnamefont
  {Davoudi}}\ and\ \bibinfo {author} {\bibfnamefont {A.-M.~S.}\ \bibnamefont
  {Tremblay}},\ }\bibfield  {title} {\bibinfo {title} {Nearest-neighbor
  repulsion and competing charge and spin order in the extended {H}ubbard
  model},\ }\href {https://doi.org/10.1103/PhysRevB.74.035113} {\bibfield
  {journal} {\bibinfo  {journal} {Phys. Rev. B}\ }\textbf {\bibinfo {volume}
  {74}},\ \bibinfo {pages} {035113} (\bibinfo {year} {2006})}\BibitemShut
  {NoStop}%
\bibitem [{\citenamefont {Paki}\ \emph {et~al.}(2019)\citenamefont {Paki},
  \citenamefont {Terletska}, \citenamefont {Iskakov},\ and\ \citenamefont
  {Gull}}]{Paki2019}%
  \BibitemOpen
  \bibfield  {author} {\bibinfo {author} {\bibfnamefont {J.}~\bibnamefont
  {Paki}}, \bibinfo {author} {\bibfnamefont {H.}~\bibnamefont {Terletska}},
  \bibinfo {author} {\bibfnamefont {S.}~\bibnamefont {Iskakov}},\ and\ \bibinfo
  {author} {\bibfnamefont {E.}~\bibnamefont {Gull}},\ }\bibfield  {title}
  {\bibinfo {title} {Charge order and antiferromagnetism in the extended
  {H}ubbard model},\ }\href {https://doi.org/10.1103/PhysRevB.99.245146}
  {\bibfield  {journal} {\bibinfo  {journal} {Phys. Rev. B}\ }\textbf {\bibinfo
  {volume} {99}},\ \bibinfo {pages} {245146} (\bibinfo {year}
  {2019})}\BibitemShut {NoStop}%
\bibitem [{\citenamefont {Huang}\ \emph {et~al.}(2013)\citenamefont {Huang},
  \citenamefont {Lai}, \citenamefont {Shi},\ and\ \citenamefont
  {Tsai}}]{Huang2013}%
  \BibitemOpen
  \bibfield  {author} {\bibinfo {author} {\bibfnamefont {W.-M.}\ \bibnamefont
  {Huang}}, \bibinfo {author} {\bibfnamefont {C.-Y.}\ \bibnamefont {Lai}},
  \bibinfo {author} {\bibfnamefont {C.}~\bibnamefont {Shi}},\ and\ \bibinfo
  {author} {\bibfnamefont {S.-W.}\ \bibnamefont {Tsai}},\ }\bibfield  {title}
  {\bibinfo {title} {Unconventional superconducting phases for the
  two-dimensional extended {H}ubbard model on a square lattice},\ }\href
  {https://doi.org/10.1103/PhysRevB.88.054504} {\bibfield  {journal} {\bibinfo
  {journal} {Phys. Rev. B}\ }\textbf {\bibinfo {volume} {88}},\ \bibinfo
  {pages} {054504} (\bibinfo {year} {2013})}\BibitemShut {NoStop}%
\bibitem [{\citenamefont {Reymbaut}\ \emph {et~al.}(2016)\citenamefont
  {Reymbaut}, \citenamefont {Charlebois}, \citenamefont {Asiani}, \citenamefont
  {1Fratino}, \citenamefont {S\'emon}, \citenamefont {Sordi},\ and\
  \citenamefont {Tremblay}}]{Reymbaut2016}%
  \BibitemOpen
  \bibfield  {author} {\bibinfo {author} {\bibfnamefont {A.}~\bibnamefont
  {Reymbaut}}, \bibinfo {author} {\bibfnamefont {M.}~\bibnamefont
  {Charlebois}}, \bibinfo {author} {\bibfnamefont {M.~F.}\ \bibnamefont
  {Asiani}}, \bibinfo {author} {\bibfnamefont {L.}~\bibnamefont {1Fratino}},
  \bibinfo {author} {\bibfnamefont {P.}~\bibnamefont {S\'emon}}, \bibinfo
  {author} {\bibfnamefont {G.}~\bibnamefont {Sordi}},\ and\ \bibinfo {author}
  {\bibfnamefont {A.-M.~S.}\ \bibnamefont {Tremblay}},\ }\bibfield  {title}
  {\bibinfo {title} {Antagonistic effects of nearest-neighbor repulsion on the
  superconducting pairing dynamics in the doped {M}ott insulator regime},\
  }\href {https://doi.org/10.1103/PhysRevB.94.155146} {\bibfield  {journal}
  {\bibinfo  {journal} {Phys. Rev. B}\ }\textbf {\bibinfo {volume} {94}},\
  \bibinfo {pages} {155146} (\bibinfo {year} {2016})}\BibitemShut {NoStop}%
\bibitem [{\citenamefont {Jiang}\ \emph {et~al.}(2018)\citenamefont {Jiang},
  \citenamefont {H\"ahner}, \citenamefont {Schulthess},\ and\ \citenamefont
  {Maier}}]{Jiang2018}%
  \BibitemOpen
  \bibfield  {author} {\bibinfo {author} {\bibfnamefont {M.}~\bibnamefont
  {Jiang}}, \bibinfo {author} {\bibfnamefont {U.~R.}\ \bibnamefont {H\"ahner}},
  \bibinfo {author} {\bibfnamefont {T.~C.}\ \bibnamefont {Schulthess}},\ and\
  \bibinfo {author} {\bibfnamefont {T.~A.}\ \bibnamefont {Maier}},\ }\bibfield
  {title} {\bibinfo {title} {d-wave superconductivity in the
  presence of nearest-neighbor {C}oulomb repulsion},\ }\href
  {https://doi.org/10.1103/PhysRevB.97.184507} {\bibfield  {journal} {\bibinfo
  {journal} {Phys. Rev. B}\ }\textbf {\bibinfo {volume} {97}},\ \bibinfo
  {pages} {184507} (\bibinfo {year} {2018})}\BibitemShut {NoStop}%
\bibitem [{\citenamefont {Chen}\ \emph {et~al.}(2023)\citenamefont {Chen},
  \citenamefont {Wang},\ and\ \citenamefont {Chen}}]{Chen2023}%
  \BibitemOpen
  \bibfield  {author} {\bibinfo {author} {\bibfnamefont {W.-C.}\ \bibnamefont
  {Chen}}, \bibinfo {author} {\bibfnamefont {Y.}~\bibnamefont {Wang}},\ and\
  \bibinfo {author} {\bibfnamefont {C.-C.}\ \bibnamefont {Chen}},\ }\bibfield
  {title} {\bibinfo {title} {Superconducting phases of the square-lattice
  extended {H}ubbard model},\ }\href
  {https://doi.org/10.1103/PhysRevB.108.064514} {\bibfield  {journal} {\bibinfo
   {journal} {Phys. Rev. B}\ }\textbf {\bibinfo {volume} {108}},\ \bibinfo
  {pages} {064514} (\bibinfo {year} {2023})}\BibitemShut {NoStop}%
\bibitem [{\citenamefont {Zhang}\ \emph {et~al.}(1995)\citenamefont {Zhang},
  \citenamefont {Carlson},\ and\ \citenamefont {Gubernatis}}]{Zhang1995}%
  \BibitemOpen
  \bibfield  {author} {\bibinfo {author} {\bibfnamefont {S.}~\bibnamefont
  {Zhang}}, \bibinfo {author} {\bibfnamefont {J.}~\bibnamefont {Carlson}},\
  and\ \bibinfo {author} {\bibfnamefont {J.~E.}\ \bibnamefont {Gubernatis}},\
  }\bibfield  {title} {\bibinfo {title} {Constrained path quantum {M}onte
  {C}arlo method for fermion ground states},\ }\href
  {https://doi.org/10.1103/PhysRevLett.74.3652} {\bibfield  {journal} {\bibinfo
   {journal} {Phys. Rev. Lett.}\ }\textbf {\bibinfo {volume} {74}},\ \bibinfo
  {pages} {3652} (\bibinfo {year} {1995})}\BibitemShut {NoStop}%
\bibitem [{\citenamefont {Zhang}\ \emph {et~al.}(1997)\citenamefont {Zhang},
  \citenamefont {Carlson},\ and\ \citenamefont {Gubernatis}}]{Zhang1997}%
  \BibitemOpen
  \bibfield  {author} {\bibinfo {author} {\bibfnamefont {S.}~\bibnamefont
  {Zhang}}, \bibinfo {author} {\bibfnamefont {J.}~\bibnamefont {Carlson}},\
  and\ \bibinfo {author} {\bibfnamefont {J.~E.}\ \bibnamefont {Gubernatis}},\
  }\bibfield  {title} {\bibinfo {title} {Constrained path {M}onte {C}arlo
  method for fermion ground states},\ }\href
  {https://doi.org/10.1103/PhysRevB.55.7464} {\bibfield  {journal} {\bibinfo
  {journal} {Phys. Rev. B}\ }\textbf {\bibinfo {volume} {55}},\ \bibinfo
  {pages} {7464} (\bibinfo {year} {1997})}\BibitemShut {NoStop}%
\bibitem [{\citenamefont {White}(1992)}]{White1992}%
  \BibitemOpen
  \bibfield  {author} {\bibinfo {author} {\bibfnamefont {S.~R.}\ \bibnamefont
  {White}},\ }\bibfield  {title} {\bibinfo {title} {Density matrix formulation
  for quantum renormalization groups},\ }\href
  {https://doi.org/10.1103/PhysRevLett.69.2863} {\bibfield  {journal} {\bibinfo
   {journal} {Phys. Rev. Lett.}\ }\textbf {\bibinfo {volume} {69}},\ \bibinfo
  {pages} {2863} (\bibinfo {year} {1992})}\BibitemShut {NoStop}%
\bibitem [{\citenamefont {White}(1993)}]{White1993}%
  \BibitemOpen
  \bibfield  {author} {\bibinfo {author} {\bibfnamefont {S.~R.}\ \bibnamefont
  {White}},\ }\bibfield  {title} {\bibinfo {title} {Density-matrix algorithms
  for quantum renormalization groups},\ }\href
  {https://doi.org/10.1103/PhysRevB.48.10345} {\bibfield  {journal} {\bibinfo
  {journal} {Phys. Rev. B}\ }\textbf {\bibinfo {volume} {48}},\ \bibinfo
  {pages} {10345} (\bibinfo {year} {1993})}\BibitemShut {NoStop}%
\bibitem [{\citenamefont {Schollw\"ock}(2005)}]{Schollwock2005}%
  \BibitemOpen
  \bibfield  {author} {\bibinfo {author} {\bibfnamefont {U.}~\bibnamefont
  {Schollw\"ock}},\ }\bibfield  {title} {\bibinfo {title} {The density-matrix
  renormalization group},\ }\href {https://doi.org/10.1103/RevModPhys.77.259}
  {\bibfield  {journal} {\bibinfo  {journal} {Rev. Mod. Phys.}\ }\textbf
  {\bibinfo {volume} {77}},\ \bibinfo {pages} {259} (\bibinfo {year}
  {2005})}\BibitemShut {NoStop}%
\bibitem [{\citenamefont {Xu}\ \emph {et~al.}(2022)\citenamefont {Xu},
  \citenamefont {Shi}, \citenamefont {Vitali}, \citenamefont {Qin},\ and\
  \citenamefont {Zhang}}]{Xu2022}%
  \BibitemOpen
  \bibfield  {author} {\bibinfo {author} {\bibfnamefont {H.}~\bibnamefont
  {Xu}}, \bibinfo {author} {\bibfnamefont {H.}~\bibnamefont {Shi}}, \bibinfo
  {author} {\bibfnamefont {E.}~\bibnamefont {Vitali}}, \bibinfo {author}
  {\bibfnamefont {M.}~\bibnamefont {Qin}},\ and\ \bibinfo {author}
  {\bibfnamefont {S.}~\bibnamefont {Zhang}},\ }\bibfield  {title} {\bibinfo
  {title} {Stripes and spin-density waves in the doped two-dimensional
  {H}ubbard model: {G}round state phase diagram},\ }\href
  {https://doi.org/10.1103/PhysRevResearch.4.013239} {\bibfield  {journal}
  {\bibinfo  {journal} {Phys. Rev. Res.}\ }\textbf {\bibinfo {volume} {4}},\
  \bibinfo {pages} {013239} (\bibinfo {year} {2022})}\BibitemShut {NoStop}%
\bibitem [{\citenamefont {LeBlanc}\ \emph {et~al.}(2015)\citenamefont
  {LeBlanc}, \citenamefont {Antipov}, \citenamefont {Becca}, \citenamefont
  {Bulik}, \citenamefont {Chan}, \citenamefont {Chung}, \citenamefont {Deng},
  \citenamefont {Ferrero}, \citenamefont {Henderson}, \citenamefont
  {Jim\'enez-Hoyos}, \citenamefont {Kozik}, \citenamefont {Liu}, \citenamefont
  {Millis}, \citenamefont {Prokof'ev}, \citenamefont {Qin}, \citenamefont
  {Scuseria}, \citenamefont {Shi}, \citenamefont {Svistunov}, \citenamefont
  {Tocchio}, \citenamefont {Tupitsyn}, \citenamefont {White}, \citenamefont
  {Zhang}, \citenamefont {Zheng}, \citenamefont {Zhu},\ and\ \citenamefont
  {Gull}}]{LeBlanc2015}%
  \BibitemOpen
  \bibfield  {author} {\bibinfo {author} {\bibfnamefont {J.~P.~F.}\
  \bibnamefont {LeBlanc}}, \bibinfo {author} {\bibfnamefont {A.~E.}\
  \bibnamefont {Antipov}}, \bibinfo {author} {\bibfnamefont {F.}~\bibnamefont
  {Becca}}, \bibinfo {author} {\bibfnamefont {I.~W.}\ \bibnamefont {Bulik}},
  \bibinfo {author} {\bibfnamefont {G.~K.-L.}\ \bibnamefont {Chan}}, \bibinfo
  {author} {\bibfnamefont {C.-M.}\ \bibnamefont {Chung}}, \bibinfo {author}
  {\bibfnamefont {Y.}~\bibnamefont {Deng}}, \bibinfo {author} {\bibfnamefont
  {M.}~\bibnamefont {Ferrero}}, \bibinfo {author} {\bibfnamefont {T.~M.}\
  \bibnamefont {Henderson}}, \bibinfo {author} {\bibfnamefont {C.~A.}\
  \bibnamefont {Jim\'enez-Hoyos}}, \bibinfo {author} {\bibfnamefont
  {E.}~\bibnamefont {Kozik}}, \bibinfo {author} {\bibfnamefont {X.-W.}\
  \bibnamefont {Liu}}, \bibinfo {author} {\bibfnamefont {A.~J.}\ \bibnamefont
  {Millis}}, \bibinfo {author} {\bibfnamefont {N.~V.}\ \bibnamefont
  {Prokof'ev}}, \bibinfo {author} {\bibfnamefont {M.}~\bibnamefont {Qin}},
  \bibinfo {author} {\bibfnamefont {G.~E.}\ \bibnamefont {Scuseria}}, \bibinfo
  {author} {\bibfnamefont {H.}~\bibnamefont {Shi}}, \bibinfo {author}
  {\bibfnamefont {B.~V.}\ \bibnamefont {Svistunov}}, \bibinfo {author}
  {\bibfnamefont {L.~F.}\ \bibnamefont {Tocchio}}, \bibinfo {author}
  {\bibfnamefont {I.~S.}\ \bibnamefont {Tupitsyn}}, \bibinfo {author}
  {\bibfnamefont {S.~R.}\ \bibnamefont {White}}, \bibinfo {author}
  {\bibfnamefont {S.}~\bibnamefont {Zhang}}, \bibinfo {author} {\bibfnamefont
  {B.-X.}\ \bibnamefont {Zheng}}, \bibinfo {author} {\bibfnamefont
  {Z.}~\bibnamefont {Zhu}},\ and\ \bibinfo {author} {\bibfnamefont
  {E.}~\bibnamefont {Gull}} (\bibinfo {collaboration} {Simons Collaboration on
  the Many-Electron Problem}),\ }\bibfield  {title} {\bibinfo {title}
  {Solutions of the two-dimensional {H}ubbard model: Benchmarks and results
  from a wide range of numerical algorithms},\ }\href
  {https://doi.org/10.1103/PhysRevX.5.041041} {\bibfield  {journal} {\bibinfo
  {journal} {Phys. Rev. X}\ }\textbf {\bibinfo {volume} {5}},\ \bibinfo {pages}
  {041041} (\bibinfo {year} {2015})}\BibitemShut {NoStop}%
\bibitem [{\citenamefont {Zheng}\ \emph {et~al.}(2017)\citenamefont {Zheng},
  \citenamefont {Chung}, \citenamefont {Corboz}, \citenamefont {Ehlers},
  \citenamefont {Qin}, \citenamefont {Noack}, \citenamefont {Shi},
  \citenamefont {White}, \citenamefont {Zhang},\ and\ \citenamefont
  {Chan}}]{Zheng2017}%
  \BibitemOpen
  \bibfield  {author} {\bibinfo {author} {\bibfnamefont {B.-X.}\ \bibnamefont
  {Zheng}}, \bibinfo {author} {\bibfnamefont {C.-M.}\ \bibnamefont {Chung}},
  \bibinfo {author} {\bibfnamefont {P.}~\bibnamefont {Corboz}}, \bibinfo
  {author} {\bibfnamefont {G.}~\bibnamefont {Ehlers}}, \bibinfo {author}
  {\bibfnamefont {M.-P.}\ \bibnamefont {Qin}}, \bibinfo {author} {\bibfnamefont
  {R.~M.}\ \bibnamefont {Noack}}, \bibinfo {author} {\bibfnamefont
  {H.}~\bibnamefont {Shi}}, \bibinfo {author} {\bibfnamefont {S.~R.}\
  \bibnamefont {White}}, \bibinfo {author} {\bibfnamefont {S.}~\bibnamefont
  {Zhang}},\ and\ \bibinfo {author} {\bibfnamefont {G.~K.-L.}\ \bibnamefont
  {Chan}},\ }\bibfield  {title} {\bibinfo {title} {Stripe order in the
  underdoped region of the two-dimensional {H}ubbard model},\ }\href
  {https://doi.org/10.1126/science.aam7127} {\bibfield  {journal} {\bibinfo
  {journal} {Science}\ }\textbf {\bibinfo {volume} {358}},\ \bibinfo {pages}
  {1155} (\bibinfo {year} {2017})}\BibitemShut {NoStop}%
\bibitem [{\citenamefont {Qin}\ \emph {et~al.}(2016)\citenamefont {Qin},
  \citenamefont {Shi},\ and\ \citenamefont {Zhang}}]{Qin2016}%
  \BibitemOpen
  \bibfield  {author} {\bibinfo {author} {\bibfnamefont {M.}~\bibnamefont
  {Qin}}, \bibinfo {author} {\bibfnamefont {H.}~\bibnamefont {Shi}},\ and\
  \bibinfo {author} {\bibfnamefont {S.}~\bibnamefont {Zhang}},\ }\bibfield
  {title} {\bibinfo {title} {Coupling quantum {M}onte {C}arlo and
  independent-particle calculations: Self-consistent constraint for the sign
  problem based on the density or the density matrix},\ }\href
  {https://doi.org/10.1103/PhysRevB.94.235119} {\bibfield  {journal} {\bibinfo
  {journal} {Phys. Rev. B}\ }\textbf {\bibinfo {volume} {94}},\ \bibinfo
  {pages} {235119} (\bibinfo {year} {2016})}\BibitemShut {NoStop}%
\bibitem [{\citenamefont {Feng}\ \emph {et~al.}(2023)\citenamefont {Feng},
  \citenamefont {Ibarra-Garc\'{\i}a-Padilla}, \citenamefont {Hazzard},
  \citenamefont {Scalettar}, \citenamefont {Zhang},\ and\ \citenamefont
  {Vitali}}]{Feng2023}%
  \BibitemOpen
  \bibfield  {author} {\bibinfo {author} {\bibfnamefont {C.}~\bibnamefont
  {Feng}}, \bibinfo {author} {\bibfnamefont {E.}~\bibnamefont
  {Ibarra-Garc\'{\i}a-Padilla}}, \bibinfo {author} {\bibfnamefont {K.~R.~A.}\
  \bibnamefont {Hazzard}}, \bibinfo {author} {\bibfnamefont {R.}~\bibnamefont
  {Scalettar}}, \bibinfo {author} {\bibfnamefont {S.}~\bibnamefont {Zhang}},\
  and\ \bibinfo {author} {\bibfnamefont {E.}~\bibnamefont {Vitali}},\
  }\bibfield  {title} {\bibinfo {title} {Metal-insulator transition and quantum
  magnetism in the {SU}(3) fermi-{H}ubbard model},\ }\href
  {https://doi.org/10.1103/PhysRevResearch.5.043267} {\bibfield  {journal}
  {\bibinfo  {journal} {Phys. Rev. Res.}\ }\textbf {\bibinfo {volume} {5}},\
  \bibinfo {pages} {043267} (\bibinfo {year} {2023})}\BibitemShut {NoStop}%
\bibitem [{\citenamefont {Zwierlein}\ \emph {et~al.}(2006)\citenamefont
  {Zwierlein}, \citenamefont {Schirotzek}, \citenamefont {Schunck},\ and\
  \citenamefont {Ketterle}}]{Zwierlein2006}%
  \BibitemOpen
  \bibfield  {author} {\bibinfo {author} {\bibfnamefont {M.~W.}\ \bibnamefont
  {Zwierlein}}, \bibinfo {author} {\bibfnamefont {A.}~\bibnamefont
  {Schirotzek}}, \bibinfo {author} {\bibfnamefont {C.~H.}\ \bibnamefont
  {Schunck}},\ and\ \bibinfo {author} {\bibfnamefont {W.}~\bibnamefont
  {Ketterle}},\ }\bibfield  {title} {\bibinfo {title} {{Fermionic superfluidity
  with imbalanced spin populations}},\ }\href
  {https://doi.org/10.1126/science.1122318} {\bibfield  {journal} {\bibinfo
  {journal} {Science}\ }\textbf {\bibinfo {volume} {311}},\ \bibinfo {pages}
  {492} (\bibinfo {year} {2006})}\BibitemShut {NoStop}%
\bibitem [{\citenamefont {Partridge}\ \emph {et~al.}(2006)\citenamefont
  {Partridge}, \citenamefont {Li}, \citenamefont {Kamar}, \citenamefont
  {Liao},\ and\ \citenamefont {Hulet}}]{Partridge2006a}%
  \BibitemOpen
  \bibfield  {author} {\bibinfo {author} {\bibfnamefont {G.~B.}\ \bibnamefont
  {Partridge}}, \bibinfo {author} {\bibfnamefont {W.}~\bibnamefont {Li}},
  \bibinfo {author} {\bibfnamefont {R.~I.}\ \bibnamefont {Kamar}}, \bibinfo
  {author} {\bibfnamefont {Y.}~\bibnamefont {Liao}},\ and\ \bibinfo {author}
  {\bibfnamefont {R.~G.}\ \bibnamefont {Hulet}},\ }\bibfield  {title} {\bibinfo
  {title} {{Pairing and Phase Separation in a Polarized {F}ermi Gas}},\ }\href
  {https://www.science.org/doi/10.1126/science.1122876} {\bibfield  {journal}
  {\bibinfo  {journal} {Science}\ }\textbf {\bibinfo {volume} {311}},\ \bibinfo
  {pages} {503} (\bibinfo {year} {2006})}\BibitemShut {NoStop}%
\bibitem [{\citenamefont {Shin}\ \emph {et~al.}(2006)\citenamefont {Shin},
  \citenamefont {Zwierlein}, \citenamefont {Schunck}, \citenamefont
  {Schirotzek},\ and\ \citenamefont {Ketterle}}]{Shin2006}%
  \BibitemOpen
  \bibfield  {author} {\bibinfo {author} {\bibfnamefont {Y.}~\bibnamefont
  {Shin}}, \bibinfo {author} {\bibfnamefont {M.~W.}\ \bibnamefont {Zwierlein}},
  \bibinfo {author} {\bibfnamefont {C.~H.}\ \bibnamefont {Schunck}}, \bibinfo
  {author} {\bibfnamefont {A.}~\bibnamefont {Schirotzek}},\ and\ \bibinfo
  {author} {\bibfnamefont {W.}~\bibnamefont {Ketterle}},\ }\bibfield  {title}
  {\bibinfo {title} {{Observation of Phase Separation in a Strongly Interacting
  Imbalanced {F}ermi Gas}},\ }\href
  {https://doi.org/10.1103/PhysRevLett.97.030401} {\bibfield  {journal}
  {\bibinfo  {journal} {Phys. Rev. Lett.}\ }\textbf {\bibinfo {volume} {97}},\
  \bibinfo {pages} {030401} (\bibinfo {year} {2006})}\BibitemShut {NoStop}%
\bibitem [{\citenamefont {Schunck}\ \emph {et~al.}(2007)\citenamefont
  {Schunck}, \citenamefont {Shin}, \citenamefont {Schirotzek}, \citenamefont
  {Zwierlein},\ and\ \citenamefont {Ketterle}}]{Schunck2007a}%
  \BibitemOpen
  \bibfield  {author} {\bibinfo {author} {\bibfnamefont {C.}~\bibnamefont
  {Schunck}}, \bibinfo {author} {\bibfnamefont {Y.}~\bibnamefont {Shin}},
  \bibinfo {author} {\bibfnamefont {A.}~\bibnamefont {Schirotzek}}, \bibinfo
  {author} {\bibfnamefont {M.}~\bibnamefont {Zwierlein}},\ and\ \bibinfo
  {author} {\bibfnamefont {W.}~\bibnamefont {Ketterle}},\ }\bibfield  {title}
  {\bibinfo {title} {{Pairing without superfluidity: The ground state of an
  imbalanced fermi mixture}},\ }\bibfield  {journal} {\bibinfo  {journal}
  {Science}\ }\textbf {\bibinfo {volume} {316}},\ \href
  {https://doi.org/10.1126/science.1140749} {10.1126/science.1140749} (\bibinfo
  {year} {2007})\BibitemShut {NoStop}%
\bibitem [{\citenamefont {Shin}\ \emph {et~al.}(2008)\citenamefont {Shin},
  \citenamefont {Schunck}, \citenamefont {Schirotzek},\ and\ \citenamefont
  {Ketterle}}]{Shin2008}%
  \BibitemOpen
  \bibfield  {author} {\bibinfo {author} {\bibfnamefont {Y.-i.}\ \bibnamefont
  {Shin}}, \bibinfo {author} {\bibfnamefont {C.~H.}\ \bibnamefont {Schunck}},
  \bibinfo {author} {\bibfnamefont {A.}~\bibnamefont {Schirotzek}},\ and\
  \bibinfo {author} {\bibfnamefont {W.}~\bibnamefont {Ketterle}},\ }\bibfield
  {title} {\bibinfo {title} {{Phase diagram of a two-component Fermi gas with
  resonant interactions}},\ }\href {http://dx.doi.org/10.1038/nature06473
  http://www.nature.com/nature/journal/v451/n7179/suppinfo/nature06473{\_}S1.html}
  {\bibfield  {journal} {\bibinfo  {journal} {Nature}\ }\textbf {\bibinfo
  {volume} {451}},\ \bibinfo {pages} {689} (\bibinfo {year}
  {2008})}\BibitemShut {NoStop}%
\bibitem [{\citenamefont {Mitra}\ \emph {et~al.}(2016)\citenamefont {Mitra},
  \citenamefont {Brown}, \citenamefont {Schau\ss{}}, \citenamefont {Kondov},\
  and\ \citenamefont {Bakr}}]{Mitra2016}%
  \BibitemOpen
  \bibfield  {author} {\bibinfo {author} {\bibfnamefont {D.}~\bibnamefont
  {Mitra}}, \bibinfo {author} {\bibfnamefont {P.~T.}\ \bibnamefont {Brown}},
  \bibinfo {author} {\bibfnamefont {P.}~\bibnamefont {Schau\ss{}}}, \bibinfo
  {author} {\bibfnamefont {S.~S.}\ \bibnamefont {Kondov}},\ and\ \bibinfo
  {author} {\bibfnamefont {W.~S.}\ \bibnamefont {Bakr}},\ }\bibfield  {title}
  {\bibinfo {title} {{Phase Separation and Pair Condensation in a
  Spin-Imbalanced 2D {F}ermi Gas}},\ }\href
  {https://doi.org/10.1103/PhysRevLett.117.093601} {\bibfield  {journal}
  {\bibinfo  {journal} {Phys. Rev. Lett.}\ }\textbf {\bibinfo {volume} {117}},\
  \bibinfo {pages} {093601} (\bibinfo {year} {2016})}\BibitemShut {NoStop}%
\bibitem [{\citenamefont {Schirotzek}\ \emph {et~al.}(2009)\citenamefont
  {Schirotzek}, \citenamefont {Wu}, \citenamefont {Sommer},\ and\ \citenamefont
  {Zwierlein}}]{Schirotzek2009}%
  \BibitemOpen
  \bibfield  {author} {\bibinfo {author} {\bibfnamefont {A.}~\bibnamefont
  {Schirotzek}}, \bibinfo {author} {\bibfnamefont {C.-H.}\ \bibnamefont {Wu}},
  \bibinfo {author} {\bibfnamefont {A.}~\bibnamefont {Sommer}},\ and\ \bibinfo
  {author} {\bibfnamefont {M.}~\bibnamefont {Zwierlein}},\ }\bibfield  {title}
  {\bibinfo {title} {{Observation of {F}ermi polarons in a tunable {F}ermi
  liquid of ultracold atoms}},\ }\bibfield  {journal} {\bibinfo  {journal}
  {Physical Review Letters}\ }\textbf {\bibinfo {volume} {102}},\ \href
  {https://doi.org/10.1103/PhysRevLett.102.230402}
  {10.1103/PhysRevLett.102.230402} (\bibinfo {year} {2009})\BibitemShut
  {NoStop}%
\bibitem [{\citenamefont {Liao}\ \emph {et~al.}(2010)\citenamefont {Liao},
  \citenamefont {Rittner}, \citenamefont {Paprotta}, \citenamefont {Li},
  \citenamefont {Partridge}, \citenamefont {Hulet}, \citenamefont {Baur},\ and\
  \citenamefont {Mueller}}]{Liao2010}%
  \BibitemOpen
  \bibfield  {author} {\bibinfo {author} {\bibfnamefont {Y.-a.}\ \bibnamefont
  {Liao}}, \bibinfo {author} {\bibfnamefont {A.~S.~C.}\ \bibnamefont
  {Rittner}}, \bibinfo {author} {\bibfnamefont {T.}~\bibnamefont {Paprotta}},
  \bibinfo {author} {\bibfnamefont {W.}~\bibnamefont {Li}}, \bibinfo {author}
  {\bibfnamefont {G.~B.}\ \bibnamefont {Partridge}}, \bibinfo {author}
  {\bibfnamefont {R.~G.}\ \bibnamefont {Hulet}}, \bibinfo {author}
  {\bibfnamefont {S.~K.}\ \bibnamefont {Baur}},\ and\ \bibinfo {author}
  {\bibfnamefont {E.~J.}\ \bibnamefont {Mueller}},\ }\bibfield  {title}
  {\bibinfo {title} {{Spin-imbalance in a one-dimensional {F}ermi gas}},\
  }\href {https://doi.org/10.1038/nature09393} {\bibfield  {journal} {\bibinfo
  {journal} {Nature}\ }\textbf {\bibinfo {volume} {467}},\ \bibinfo {pages}
  {567} (\bibinfo {year} {2010})}\BibitemShut {NoStop}%
\bibitem [{\citenamefont {Massignan}\ \emph {et~al.}(2025)\citenamefont
  {Massignan}, \citenamefont {Schmidt}, \citenamefont {Astrakharchik},
  \citenamefont {İmamoglu}, \citenamefont {Zwierlein}, \citenamefont {Arlt},\
  and\ \citenamefont {Bruun}}]{Massignan2025}%
  \BibitemOpen
  \bibfield  {author} {\bibinfo {author} {\bibfnamefont {P.}~\bibnamefont
  {Massignan}}, \bibinfo {author} {\bibfnamefont {R.}~\bibnamefont {Schmidt}},
  \bibinfo {author} {\bibfnamefont {G.~E.}\ \bibnamefont {Astrakharchik}},
  \bibinfo {author} {\bibfnamefont {A.}~\bibnamefont {İmamoglu}}, \bibinfo
  {author} {\bibfnamefont {M.}~\bibnamefont {Zwierlein}}, \bibinfo {author}
  {\bibfnamefont {J.~J.}\ \bibnamefont {Arlt}},\ and\ \bibinfo {author}
  {\bibfnamefont {G.~M.}\ \bibnamefont {Bruun}},\ }\href
  {https://arxiv.org/abs/2501.09618} {\bibinfo {title} {{Polarons in atomic
  gases and two-dimensional semiconductors}}} (\bibinfo {year} {2025}),\
  \Eprint {https://arxiv.org/abs/2501.09618} {arXiv:2501.09618
  [cond-mat.quant-gas]} \BibitemShut {NoStop}%
\bibitem [{\citenamefont {Feng}\ \emph {et~al.}(2025)\citenamefont {Feng},
  \citenamefont {Hartke}, \citenamefont {He}, \citenamefont {Oreg},
  \citenamefont {Turnbaugh}, \citenamefont {Jia}, \citenamefont {Zwierlein},\
  and\ \citenamefont {Zhang}}]{Feng2025}%
  \BibitemOpen
  \bibfield  {author} {\bibinfo {author} {\bibfnamefont {C.}~\bibnamefont
  {Feng}}, \bibinfo {author} {\bibfnamefont {T.}~\bibnamefont {Hartke}},
  \bibinfo {author} {\bibfnamefont {Y.-Y.}\ \bibnamefont {He}}, \bibinfo
  {author} {\bibfnamefont {B.}~\bibnamefont {Oreg}}, \bibinfo {author}
  {\bibfnamefont {C.}~\bibnamefont {Turnbaugh}}, \bibinfo {author}
  {\bibfnamefont {N.}~\bibnamefont {Jia}}, \bibinfo {author} {\bibfnamefont
  {M.}~\bibnamefont {Zwierlein}},\ and\ \bibinfo {author} {\bibfnamefont
  {S.}~\bibnamefont {Zhang}},\ }\href {https://arxiv.org/abs/2509.02688}
  {\bibinfo {title} {In search of exotic pairing in the {H}ubbard model:
  many-body computation and quantum gas microscopy}} (\bibinfo {year} {2025}),\
  \Eprint {https://arxiv.org/abs/2509.02688} {arXiv:2509.02688
  [cond-mat.quant-gas]} \BibitemShut {NoStop}%
\bibitem [{\citenamefont {Hartke}\ \emph {et~al.}(2025)\citenamefont {Hartke},
  \citenamefont {Oreg}, \citenamefont {Feng}, \citenamefont {Turnbaugh},
  \citenamefont {Hertkorn}, \citenamefont {He}, \citenamefont {Jia},
  \citenamefont {Khatami}, \citenamefont {Zhang},\ and\ \citenamefont
  {Zwierlein}}]{Hartke2025}%
  \BibitemOpen
  \bibfield  {author} {\bibinfo {author} {\bibfnamefont {T.}~\bibnamefont
  {Hartke}}, \bibinfo {author} {\bibfnamefont {B.}~\bibnamefont {Oreg}},
  \bibinfo {author} {\bibfnamefont {C.}~\bibnamefont {Feng}}, \bibinfo {author}
  {\bibfnamefont {C.}~\bibnamefont {Turnbaugh}}, \bibinfo {author}
  {\bibfnamefont {J.}~\bibnamefont {Hertkorn}}, \bibinfo {author}
  {\bibfnamefont {Y.-Y.}\ \bibnamefont {He}}, \bibinfo {author} {\bibfnamefont
  {N.}~\bibnamefont {Jia}}, \bibinfo {author} {\bibfnamefont {E.}~\bibnamefont
  {Khatami}}, \bibinfo {author} {\bibfnamefont {S.}~\bibnamefont {Zhang}},\
  and\ \bibinfo {author} {\bibfnamefont {M.}~\bibnamefont {Zwierlein}},\ }\href
  {https://arxiv.org/abs/2511.10605} {\bibinfo {title} {Competition of fermion
  pairing, magnetism, and charge order in the spin-doped attractive {H}ubbard
  gas}} (\bibinfo {year} {2025}),\ \Eprint {https://arxiv.org/abs/2511.10605}
  {arXiv:2511.10605 [cond-mat.quant-gas]} \BibitemShut {NoStop}%
\bibitem [{\citenamefont {Xu}\ \emph {et~al.}(2025{\natexlab{a}})\citenamefont
  {Xu}, \citenamefont {Kendrick}, \citenamefont {Kale}, \citenamefont {Gang},
  \citenamefont {Feng}, \citenamefont {Zhang}, \citenamefont {Young},
  \citenamefont {Lebrat},\ and\ \citenamefont {Greiner}}]{Xu2025}%
  \BibitemOpen
  \bibfield  {author} {\bibinfo {author} {\bibfnamefont {M.}~\bibnamefont
  {Xu}}, \bibinfo {author} {\bibfnamefont {L.~H.}\ \bibnamefont {Kendrick}},
  \bibinfo {author} {\bibfnamefont {A.}~\bibnamefont {Kale}}, \bibinfo {author}
  {\bibfnamefont {Y.}~\bibnamefont {Gang}}, \bibinfo {author} {\bibfnamefont
  {C.}~\bibnamefont {Feng}}, \bibinfo {author} {\bibfnamefont {S.}~\bibnamefont
  {Zhang}}, \bibinfo {author} {\bibfnamefont {A.~W.}\ \bibnamefont {Young}},
  \bibinfo {author} {\bibfnamefont {M.}~\bibnamefont {Lebrat}},\ and\ \bibinfo
  {author} {\bibfnamefont {M.}~\bibnamefont {Greiner}},\ }\bibfield  {title}
  {\bibinfo {title} {{A neutral-atom {H}ubbard quantum simulator in the
  cryogenic regime}},\ }\href {https://doi.org/10.1038/s41586-025-09112-w}
  {\bibfield  {journal} {\bibinfo  {journal} {Nature}\ }\textbf {\bibinfo
  {volume} {642}},\ \bibinfo {pages} {909} (\bibinfo {year}
  {2025}{\natexlab{a}})}\BibitemShut {NoStop}%
\bibitem [{\citenamefont {Guardado-Sanchez}\ \emph {et~al.}(2021)\citenamefont
  {Guardado-Sanchez}, \citenamefont {Spar}, \citenamefont {Schauss},
  \citenamefont {Belyansky}, \citenamefont {Young}, \citenamefont {Bienias},
  \citenamefont {Gorshkov}, \citenamefont {Iadecola},\ and\ \citenamefont
  {Bakr}}]{Guardado-Sanchez2021}%
  \BibitemOpen
  \bibfield  {author} {\bibinfo {author} {\bibfnamefont {E.}~\bibnamefont
  {Guardado-Sanchez}}, \bibinfo {author} {\bibfnamefont {B.~M.}\ \bibnamefont
  {Spar}}, \bibinfo {author} {\bibfnamefont {P.}~\bibnamefont {Schauss}},
  \bibinfo {author} {\bibfnamefont {R.}~\bibnamefont {Belyansky}}, \bibinfo
  {author} {\bibfnamefont {J.~T.}\ \bibnamefont {Young}}, \bibinfo {author}
  {\bibfnamefont {P.}~\bibnamefont {Bienias}}, \bibinfo {author} {\bibfnamefont
  {A.~V.}\ \bibnamefont {Gorshkov}}, \bibinfo {author} {\bibfnamefont
  {T.}~\bibnamefont {Iadecola}},\ and\ \bibinfo {author} {\bibfnamefont
  {W.~S.}\ \bibnamefont {Bakr}},\ }\bibfield  {title} {\bibinfo {title} {Quench
  dynamics of a {F}ermi gas with strong nonlocal interactions},\ }\href
  {https://doi.org/10.1103/PhysRevX.11.021036} {\bibfield  {journal} {\bibinfo
  {journal} {Phys. Rev. X}\ }\textbf {\bibinfo {volume} {11}},\ \bibinfo
  {pages} {021036} (\bibinfo {year} {2021})}\BibitemShut {NoStop}%
\bibitem [{\citenamefont {Langen}\ \emph {et~al.}(2024)\citenamefont {Langen},
  \citenamefont {Valtolina}, \citenamefont {Wang},\ and\ \citenamefont
  {Ye}}]{Langen2024}%
  \BibitemOpen
  \bibfield  {author} {\bibinfo {author} {\bibfnamefont {T.}~\bibnamefont
  {Langen}}, \bibinfo {author} {\bibfnamefont {G.}~\bibnamefont {Valtolina}},
  \bibinfo {author} {\bibfnamefont {D.}~\bibnamefont {Wang}},\ and\ \bibinfo
  {author} {\bibfnamefont {J.}~\bibnamefont {Ye}},\ }\bibfield  {title}
  {\bibinfo {title} {Quantum state manipulation and cooling of ultracold
  molecules},\ }\href {https://doi.org/10.1038/s41567-024-02423-1} {\bibfield
  {journal} {\bibinfo  {journal} {Nature Physics}\ }\textbf {\bibinfo {volume}
  {20}},\ \bibinfo {pages} {702} (\bibinfo {year} {2024})}\BibitemShut
  {NoStop}%
\bibitem [{\citenamefont {Cornish}\ \emph {et~al.}(2024)\citenamefont
  {Cornish}, \citenamefont {Tarbutt},\ and\ \citenamefont
  {Hazzard}}]{Cornish2024}%
  \BibitemOpen
  \bibfield  {author} {\bibinfo {author} {\bibfnamefont {S.~L.}\ \bibnamefont
  {Cornish}}, \bibinfo {author} {\bibfnamefont {M.~R.}\ \bibnamefont
  {Tarbutt}},\ and\ \bibinfo {author} {\bibfnamefont {K.~R.~A.}\ \bibnamefont
  {Hazzard}},\ }\bibfield  {title} {\bibinfo {title} {Quantum computation and
  quantum simulation with ultracold molecules},\ }\href
  {https://doi.org/10.1038/s41567-024-02453-9} {\bibfield  {journal} {\bibinfo
  {journal} {Nature Physics}\ }\textbf {\bibinfo {volume} {20}},\ \bibinfo
  {pages} {730} (\bibinfo {year} {2024})}\BibitemShut {NoStop}%
\bibitem [{\citenamefont {Xu}\ \emph {et~al.}(2025{\natexlab{b}})\citenamefont
  {Xu}, \citenamefont {Fischer}, \citenamefont {Tancogne-Dejean}, \citenamefont
  {Zhang}, \citenamefont {Bostr\"om}, \citenamefont {Claassen}, \citenamefont
  {Kennes}, \citenamefont {Rubio},\ and\ \citenamefont
  {Xian}}]{XuQiaoling2025}%
  \BibitemOpen
  \bibfield  {author} {\bibinfo {author} {\bibfnamefont {Q.}~\bibnamefont
  {Xu}}, \bibinfo {author} {\bibfnamefont {A.}~\bibnamefont {Fischer}},
  \bibinfo {author} {\bibfnamefont {N.}~\bibnamefont {Tancogne-Dejean}},
  \bibinfo {author} {\bibfnamefont {T.}~\bibnamefont {Zhang}}, \bibinfo
  {author} {\bibfnamefont {E.~V.~n.}\ \bibnamefont {Bostr\"om}}, \bibinfo
  {author} {\bibfnamefont {M.}~\bibnamefont {Claassen}}, \bibinfo {author}
  {\bibfnamefont {D.~M.}\ \bibnamefont {Kennes}}, \bibinfo {author}
  {\bibfnamefont {A.}~\bibnamefont {Rubio}},\ and\ \bibinfo {author}
  {\bibfnamefont {L.}~\bibnamefont {Xian}},\ }\bibfield  {title} {\bibinfo
  {title} {Engineering 2d square lattice {H}ubbard models in
  90\ifmmode^\circ\else\textdegree\fi{} twisted $\mathrm{GeX}/\mathrm{SnX}$
  ($\mathrm{X}=\mathrm{S}$, se) {M}oir\'e superlattices},\ }\href
  {https://doi.org/10.1103/wcbz-lbr1} {\bibfield  {journal} {\bibinfo
  {journal} {Phys. Rev. X}\ }\textbf {\bibinfo {volume} {15}},\ \bibinfo
  {pages} {041049} (\bibinfo {year} {2025}{\natexlab{b}})}\BibitemShut
  {NoStop}%
\end{thebibliography}%

\end{document}

\preprint{}
	
	\title{Interaction-Driven Ferrimagnetic Stripes in the Extended Hubbard Model}
	
	\author{Chunhan Feng}
	\affiliation{Center for Computational Quantum Physics, Flatiron Institute, 162 5th Avenue, New York, New York, USA}
	\affiliation{Max Planck Institute for the Physics of Complex Systems, Nöthnitzer Straße 38, 01187 Dresden, Germany}
	\author{ Miguel A. Morales}
	\affiliation{Center for Computational Quantum Physics, Flatiron Institute, 162 5th Avenue, New York, New York, USA}
	\author{Shiwei Zhang}
	\affiliation{Center for Computational Quantum Physics, Flatiron Institute, 162 5th Avenue, New York, New York, USA}
	
	\date{\today}

	\clearpage
	\centerline{\bf Supplemental Materials}
	\vskip0.10in

	\vskip0.10in \noindent
	In these Supplemental Materials, we present additional details concerning: 1)computational details; 
	2) energy comparison of nearly degenerate ground states; 
	3) finite lattice size effects; 4) ferrimagnetic order.
	
	\vskip0.10in \noindent
	{\it Computational details}
	
	(a) \textbf{Pinning field.} We introduce a weak boundary pinning field $H_p=\sum_{i_y \in odd}-(-1)^{i_y+1}v_pn_{i\uparrow}$ (and $H_p=\sum_{i_y \in even}\mp (-1)^{i_y+L_x}v_pn_{i\uparrow}$) on the left edge (or both edges) with strength $v_p=0.1$. At this amplitude, the field does not affect bulk observables and merely selects among nearly degenerate symmetry-related configurations. \cite{Xu2022}

	(b) \textbf{DMRG.} In our DMRG simulations, we use a truncation error of $10^{-8}$
	and a maximum bond dimension of $\mathrm{MBD}=12000$. When the desired truncation error cannot be reached, the maximum bond dimension effectively controls the numerical accuracy.
	
	(c) \textbf{Hartree–Fock approximations.} In the unrestricted Hartree Fock (UHF) calculation, the onsite and nearest-neighbor interactions are decoupled as $ H^{\rm U}_{\rm UHF}=U_{\rm eff} 
	\sum_{i} \left(  \langle n_{i\uparrow}\rangle n_{i\downarrow}+  \langle n_{i\downarrow}\rangle n_{i\uparrow}- \langle n_{i\uparrow}\rangle \langle n_{i\downarrow} \rangle \right)$ 
	and $ H^{\rm V}_{\rm  UHF}=V_{\rm eff}\sum_{\langle i,j \rangle}(n_i \langle n_j \rangle + n_j \langle n_i \rangle - \langle n_i \rangle \langle n_j \rangle -\sum_{\sigma}
	(\langle c_{i\sigma}^\dagger c_{j\sigma} \rangle c_{j\sigma}^\dagger c_{i\sigma} 
	+\langle c_{j\sigma}^\dagger c_{i\sigma} \rangle c_{i\sigma}^\dagger c_{j\sigma} 
	-\langle c_{i\sigma}^\dagger c_{j\sigma} \rangle \langle c_{j\sigma}^\dagger c_{i\sigma} \rangle ))$. In the generalized Hartree Fock (GHF) calculation, additional spin-flip channels are included. The onsite and nearest-neighbor contributions become
	$H^{\rm U}_{\rm GHF}=- U_{\rm eff} \sum_i ( \langle c_{i\downarrow}^{\dagger} c_{i\uparrow}  \rangle c_{i\uparrow}^{\dagger} c_{i\downarrow}   
	+\langle  c_{i\uparrow}^{\dagger} c_{i\downarrow}   \rangle
	c_{i\downarrow}^{\dagger} c_{i\uparrow} 
	-\langle  c_{i\uparrow}^{\dagger} c_{i\downarrow}   \rangle
	\langle c_{i\downarrow}^{\dagger} c_{i\uparrow} \rangle
	)$ together with the corresponding spin-exchange decouplings in the $V$ term, 
	$H^{\rm V}_{\rm GHF}=- V_{\rm eff} \sum_i ( \langle c_{j\downarrow}^{\dagger} c_{i\uparrow}  \rangle c_{i\uparrow}^{\dagger} c_{j\downarrow}   
	+\langle  c_{i\uparrow}^{\dagger} c_{j\downarrow}   \rangle
	c_{j\downarrow}^{\dagger} c_{i\uparrow} 
	-\langle  c_{i\uparrow}^{\dagger} c_{j\downarrow}   \rangle
	\langle c_{j\downarrow}^{\dagger} c_{i\uparrow} 
	\rangle
	+\langle c_{i\downarrow}^{\dagger} c_{j\uparrow}  \rangle c_{j\uparrow}^{\dagger} c_{i\downarrow}   
	+\langle  c_{j\uparrow}^{\dagger} c_{i\downarrow}   \rangle
	c_{i\downarrow}^{\dagger} c_{j\uparrow} 
	-\langle  c_{j\uparrow}^{\dagger} c_{i\downarrow}   \rangle
	\langle c_{i\downarrow}^{\dagger} c_{j\uparrow} 
	\rangle
	)$ allowing for spin-imbalanced magnetic order.
	
	\begin{figure}[h]
		\includegraphics[width=1\columnwidth]{SM_diff_init_start.png}
		\caption{Spin density $S_i^z$along a line cut and the averaged hole density $n_{\rm hole}$ for $U=8, V/U=0.3$ at doping $\delta=1/6$. Red and blue curves denote the converged solutions obtained from two different initial trial wave functions. The red curves start from 
			$U_{\rm eff}=2.75, V_{\rm eff}/U_{\rm eff}=0$(light red), corresponding to a stripe phase of the pure Hubbard model. The blue curves start from $U_{\rm eff}=3, V_{\rm eff}/U_{\rm eff}=0.4$ (light blue), corresponding to an alternating spin-polarized stripe phase.
		}
		\label{fig:SM_diff_init_start}
	\end{figure}
	
	\begin{figure}[h]
		\includegraphics[width=1\columnwidth]{ABAB_AABB.png}
		\includegraphics[width=1\columnwidth]{E_ABAB_AABB.png}
		\caption{Representative configurations of two nearly degenerate polarized stripe states at $V/U \gtrsim 0.25$ with magnetization $m=0$. Upper and middle panels show spin and charge densities, respectively. 
			Left: alternating polarized spin stripes (“ABAB” pattern). Right: clustered polarized regions (“AABB” pattern).
			Lower panels: energy comparison of these two states at different dopings $\delta=1/8, 1/6$,and $1/4$ on different lattice sizes represented by different curves.
		}
		\label{fig:ABAB_AABB}
	\end{figure}
	
	(d)\textbf{Independence on initial states.} In Fig.~\ref{fig:SM_diff_init_start}, we show the spin density $S_i^z$ along a line cut (left panel) and the averaged hole densities $n_{\rm hole}$(right panel) obtained from two distinct initial trial wave functions. The light red and blue curves correspond to initial UHF solutions with $U_{\rm eff}=2.75$, $V_{\rm eff}/U_{\rm eff}=0$ (conventional spin stripe phase of the pure Hubbard model) and $U_{\rm eff}=3$, $V_{\rm eff}/U_{\rm eff}=0.4$ (polarized spin stripe phase), respectively. After two self-consistency cycles, the converged results from both initial states are nearly indistinguishable (red and blue curves). This demonstrates the robustness of the polarized spin stripe order.
	
	\begin{figure}[b]
		\includegraphics[width=1\columnwidth]{lattice_size_effects.png}
		\caption{Lattice size effects on the polarized spin stripes for $U=8, V/U=0.3$ at doping $\delta=1/6$. Left: spin density along a line cut; right: averaged hole density along the long direction. Different curves represent different lattice sizes as labeled on the top of the plot.\textcolor{blue}{$24 \times 12$ to be added.}
		}
		\label{fig:lattice_size_effects}
	\end{figure}
	\vskip0.10in \noindent
	{\it Nearly degenerate polarized spin stripe phase}
	
	In the quantum Monte Carlo simulations with fixed total magnetization $m=0$, we observe two nearly degenerate ground states when $V/U \gtrsim 0.25$. Representative configurations for a $24 \times 6$ lattice at doping $\delta=1/8$ are shown in Fig.~\ref{fig:ABAB_AABB}.
	The left panel (“ABAB”) corresponds to the alternating polarized spin stripe phase discussed in the main text, where bluish and pinkish regions alternate. The right panel (“AABB”) represents a competing state in which like-polarized regions cluster and are spatially separated. The energy difference per site is of order $10^{-4}$ in many body calculation, which is beyond the resolution of QMC calculations. In the Hartree Fock, ''AABB" usually has a lower energy compared with ''ABAB" on larger lattice sizes, reflecting a tendency toward ferrimagnetic order when the total magnetization is allowed to vary.
	\begin{figure}[t]
		\includegraphics[width=1\columnwidth]{SM_E_vs_m.png}
		\caption{ Energy $E$ as a function of magnetization $m$ for systems with doping $\delta=1/12$ (panels (a)(c)) and 
			$\delta=1/4$ (panels (b)(d)). Calculations are performed on a $24 \times 6$ lattice with $U=8$. $V$ values are labeled on the top of each panel.}
		\label{fig:SM_E_vs_m}
	\end{figure}
	
	\begin{figure}[h]
		\includegraphics[width=1\columnwidth]{SM_E_ABAB_AABB_Ferri.png}
		\caption{ Energy difference (per site) as a function of  $V/U$ for $24\times 6$ lattice at doping $\delta=1/6$ for $U=8$. ABAB state is used as a reference.
		}
		\label{fig:SM_E_ABAB_AABB_Ferri}
	\end{figure}
	\vskip0.10in \noindent
	{\it Lattice size effects}
	
	In Fig.~\ref{fig:lattice_size_effects}, we show the spin density $S$ along a line-cut and hole densities averaged over all the rows for U=8 system with hole doping $\delta=1/6$ on different lattice sizes (represented by different curves) with length up to $L_x=36$ and width up to $L_y=12$. Except the boundary, different curves are basically on top of each other, implying the lattice size effect is limited.

	\vskip0.10in \noindent
	{\it Ferrimagnetic order}
	
	In Fig.~\ref{fig:SM_E_vs_m}, we present the energy per site $E$ as a function of magnetization $m$ for a $24 \times 6$ lattice at dopings (a,c)$\delta=1/12$ and (b,d)$\delta=1/4$.
	At $V/U=0.2$, the lowest-energy state for both dopings corresponds to zero magnetization. In contrast, at $V/U=0.3$, a finite magnetization $m=\delta/2$ becomes energetically degenerate with $m=0$ state at (c) $\delta=1/12$ and  energetically  favored at (d) $\delta=1/4$, indicating a much stronger tendency toward ferrimagnetic order at larger doping. Finally we plot the energy difference per site as a function of $V/U$ for filling $\delta=1/6$ in Fig.~\ref{fig:SM_E_ABAB_AABB_Ferri}. The energy differences among the ABAB, AABB, and ferrimagnetic states are of order $10^{-4}$, indicating that they are nearly degenerate. 
		\bibliography{Extended_Hubbard_with_V}